\documentclass[fleqn,usenatbib]{mnras}

\usepackage{newtxtext,newtxmath}

\usepackage[T1]{fontenc}

\DeclareRobustCommand{\VAN}[3]{#2}
\let\VANthebibliography\thebibliography
\def\thebibliography{\DeclareRobustCommand{\VAN}[3]{##3}\VANthebibliography}

\usepackage{graphicx}	
\usepackage{subcaption} 
\usepackage{float}      
\usepackage{amsmath}	
\usepackage{kotex}      
\usepackage{comment}
\usepackage[normalem]{ulem} 
\usepackage{natbib}     

\usepackage{orcidlink}

\usepackage{anyfontsize}

\usepackage{xspace}

\usepackage{cleveref}

\newcommand{\Crefapp}[1]{Appendix~\ref{#1}}

\defcitealias{Bruzual&Charlot2003MNRAS.344.1000B}{BC03}
\defcitealias{Childress+2014MNRAS.445.1898C}{C14}
\defcitealias{Lee+2022MNRAS.517.2697L}{L22}
\defcitealias{Chung+2025MNRAS.538.3340C}{C25}
\newcommand{\PaperI}{Paper I (\citetalias{Chung+2025MNRAS.538.3340C})\xspace}
\defcitealias{Son+2025MNRAS.544..975S}{S25}
\newcommand{\PaperII}{Paper II (\citetalias{Son+2025MNRAS.544..975S})\xspace}
\defcitealias{Rose+2019ApJ...874...32R}{R19}
\defcitealias{Gupta+2011ApJ...740...92G}{G11}

\title[Age Bias in SN Cosmology -- III]{Strong Progenitor Age Bias in Supernova Cosmology. III. \\
Progenitor Age as the Physical Origin of the Type Ia Supernova \\
Magnitude Steps with Host Properties}

\author[Park et al.]
{Seunghyun Park\orcidlink{0009-0002-1351-1582},\thanks{E-mail: shpark.stellarnote@gmail.com (SHP)}
Young-Wook Lee\orcidlink{0000-0002-2210-1238},\thanks{E-mail: ywlee2@yonsei.ac.kr (Y-WL)}
Chul Chung\orcidlink{0000-0001-6812-4542},\thanks{E-mail: chulchung@yonsei.ac.kr (CC)}
Suk-Jin Yoon\orcidlink{0000-0002-1842-4325},\thanks{E-mail: sjyoon0691@yonsei.ac.kr (S-JY)}
\newauthor
Junhyuk Son\orcidlink{0009-0004-3117-1977},
Hyejeon Cho\orcidlink{0000-0001-5966-5072},
Young-Lo Kim\orcidlink{0000-0002-1031-0796}
\\
Department of Astronomy and Center for Galaxy Evolution Research, Yonsei University, Seoul 03722, Republic of Korea}

\date{Accepted XXX. Received YYY; in original form ZZZ}

\pubyear{\the\year{}}

\begin{document}
\label{firstpage}
\pagerange{\pageref{firstpage}--\pageref{lastpage}}
\maketitle







\begin{abstract}
The standardized magnitude of a type Ia supernova (SN Ia) correlates with host-galaxy properties, and a host mass-step correction is now routinely included in SN Ia luminosity standardization.
Given that host mass cannot directly influence SN Ia luminosity, the root cause of the step must be another latent parameter associated with host mass.
Identifying this driver is essential because different host properties evolve differently with redshift, so corrections based on them can lead to divergent cosmological inferences.
In recent years, direct and extensive age measurements have revealed a significant relation between host age and Hubble residual (HR).
Here, using a new dataset, we confirm that this relation arises from the age dependence of the SN Ia luminosity standardization process and the resulting overcorrection.
Specifically, we show that while the mass-step correction reduces the age bias by about half, the host age-bias correction fully eliminates the mass step, supporting a progenitor-age origin of the host-age--HR relation.
We further demonstrate that the SN Ia magnitude steps with host mass (and specific star formation rate; sSFR) emerge from a nonlinear, step-like relation between mass (and sSFR) and progenitor age, combined with a linear progenitor-age--HR relation: the SN Ia magnitude steps are therefore projected manifestations of an underlying dependence on progenitor age.
Taken together, our results show that progenitor age is the primary driver of both the strong host-age--HR relation and the apparent host-mass and host-sSFR steps.
\end{abstract}



\begin{keywords}
supernovae: general -- galaxies: evolution -- cosmology: observations
\end{keywords}



\section[Introduction]{Introduction}
\label{sec1.intro}

Type Ia supernovae (SNe Ia) constitute the upper rung of the cosmic distance ladder, extending distance measurements out to high redshift for cosmological applications.
They are ``standardizable'' candles that require a luminosity standardization process to calibrate their observed brightness \citep{Phillips1993ApJ...413L.105P, Tripp1998A&A...331..815T, Guy+2007A&A...466...11G}.
To maintain their utility as standardizable candles over cosmic time, supernova (SN) cosmology relies on the key assumption that ``the calibrating relationships between SN luminosity and light-curve shape must be invariant with progenitor age'' \citep{Jha+2019NatAs...3..706J}.
Under this fundamental assumption, the observed dimming of high redshift SNe Ia, by about 0.25~mag relative to a baseline model without dark energy, was interpreted as evidence for the accelerated expansion of the universe driven by a cosmological constant \citep{Riess+1998AJ....116.1009R, Perlmutter+1999ApJ...517..565P}.

However, numerous follow-up studies have reported that, even after luminosity standardization based on light-curve shape and color, the magnitudes of SNe Ia still depend on the properties of their host galaxies, suggesting that the above assumption may require further scrutiny.
Correlations between host galaxy morphology and standardized SN Ia magnitude had already been noted in the 1990s, and were attributed either to differences in stellar population age \citep{Hamuy+1996AJ....112.2391H, Hamuy+1996AJ....112.2398H} or to dust \citep{Sullivan+2003MNRAS.340.1057S}.
This morphology--luminosity correlation was later confirmed by \citet{Hicken+2009ApJ...700.1097H} using a substantially larger sample.
Subsequently, the discovery that SNe Ia in host galaxies with masses below $10^{10} \, \mathrm{M_{\odot}}$ are approximately 0.08~mag fainter \citep{Kelly+2010ApJ...715..743K, Sullivan+2010MNRAS.406..782S} initiated numerous efforts to verify the existence of this ``host mass step'', which has since been widely confirmed \citep[e.g.,][]{Lampeitl+2010ApJ...722..566L, Pan+2014MNRAS.438.1391P, Betoule+2014A&A...568A..22B, Campbell+2016MNRAS.457.3470C, Kim+2019JKAS...52..181K}.
Meanwhile, \citet{Rigault+2013A&A...560A..66R, Rigault+2015ApJ...802...20R, Rigault+2020A&A...644A.176R} measured the local specific star formation rate (sSFR) around SN Ia sites in host galaxies and found a significant 0.16~mag step (``sSFR step''), similar to the host mass step but of larger amplitude, which they attributed to differences in progenitor age \citep[see also][for local environment studies]{Kim+2018ApJ...854...24K, Kim+2024MNRAS.527.4359K}.

Among the various host properties correlated with standardized SN Ia magnitude, the host mass step has become a widely adopted additional correction term in the SN Ia luminosity standardization process.
However, galaxy mass itself cannot physically modify the intrinsic luminosity of a SN Ia, implying that the mass step must instead trace some underlying parameter correlated with mass---most plausibly stellar population age or metallicity \citep[see][]{Kang+2020ApJ...889....8K}.
\citet{Brout+2021ApJ...909...26B} suggested that the host-mass-dependent dust extinction parameter, the total-to-selective extinction ratio $R_V$, might be responsible for the mass step.
However, the required ad hoc mass-dependent $R_V$ values directly contradict \textit{GALEX} ultraviolet observations \citep[see Table 1 in][]{Salim+2018ApJ...859...11S}, making this explanation observationally unsubstantiated.
Moreover, dust extinction is already largely accounted for in the luminosity standardization process through the color--luminosity relation \citep{Tripp1998A&A...331..815T}.
These considerations highlight the importance of examining direct and reliable measurements of host galaxy age and metallicity to identify the true physical origin of the mass step and other host-property correlations.

Studies that directly measured host metallicity have not found statistically significant correlations \citep[see also \Cref{sec2.root} below]{Pan+2014MNRAS.438.1391P, Campbell+2016MNRAS.457.3470C, Kang+2020ApJ...889....8K}.
In contrast, direct and extensive measurements of mean stellar population ages of host galaxies (``host ages'') in recent years have revealed a strong correlation between standardized SN Ia magnitude and host age.
\citet{Kang+2020ApJ...889....8K} first reported an important hint of a $3\sigma$ correlation based on very high-quality (S/N = 175) spectra of nearby early-type host galaxies.
Subsequently, using the \citet{Rose+2019ApJ...874...32R} sample with photometrically derived ages, \citet{Lee+2020ApJ...903...22L} found a $4.3\sigma$ correlation for host galaxies comprising all morphological types.
This result was independently confirmed by two third-party teams at $\sim 5\sigma$ significance \citep{Zhang+2021MNRAS.503L..33Z, Wang+2023SCPMA..6629511W}.
In particular, Paper I of this series \citep[hereafter \citetalias{Chung+2025MNRAS.538.3340C}]{Chung+2025MNRAS.538.3340C} confirmed this correlation at $5.5\sigma$ for the low-redshift sample \citep{Rose+2019ApJ...874...32R} and demonstrated its ubiquity in an additional sample \citep{Gupta+2011ApJ...740...92G} extending out to $z \approx 0.45$.
When the three independent samples \citep{Gupta+2011ApJ...740...92G, Rose+2019ApJ...874...32R, Kang+2020ApJ...889....8K} across different redshift ranges are considered together, the combined statistical significance exceeds $7\sigma$.

Identifying the root cause of the host-property dependence is crucial because different host properties evolve very differently with redshift, and corrections based on them can lead to substantially different cosmological inferences (see \Cref{sec4.discussion} below).
The strong correlation between host age and standardized SN Ia magnitude provides compelling evidence that SN Ia progenitor age is the fundamental origin of this host-property dependence.
To investigate this in greater detail, in \Cref{sec2.root} we extend the analysis of \citet{Lee+2022MNRAS.517.2697L} to the \citet{Gupta+2011ApJ...740...92G} sample and examine the width--luminosity and color--luminosity relations in the SN Ia luminosity standardization process to assess which of the host properties---mass, stellar population age, or metallicity---most directly drives the observed host-property dependence.
In \Cref{sec3.nonlinearity}, we then use simple and intuitive simulations to show how the linear relation with progenitor age is projected into the apparent SN Ia magnitude steps with host mass and sSFR.
Finally, \Cref{sec4.discussion} discusses the cosmological implications of these findings.



\section[Identifying the Root Cause of Host-Property Dependence]{Identifying the Root Cause of Host-Property Dependence}
\label{sec2.root}



\subsection{Host-Galaxy Samples and Age Measurements}
\label{sec2.1.hostsample}

Our analysis builds on the host-galaxy samples and the updated age estimates presented in \PaperI.
In that work, we examined two SN Ia host samples based on Sloan Digital Sky Survey (SDSS) $ugriz$ photometry: \citet[hereafter \citetalias{Gupta+2011ApJ...740...92G}]{Gupta+2011ApJ...740...92G} and \citet[hereafter \citetalias{Rose+2019ApJ...874...32R}]{Rose+2019ApJ...874...32R}.
The \citetalias{Gupta+2011ApJ...740...92G} sample contains 206 hosts extending to $z \approx 0.45$, whereas \citetalias{Rose+2019ApJ...874...32R} provides a sample of 102 hosts at $z < 0.20$.
Using the latest version of the Flexible Stellar Population Synthesis (FSPS) code \citep{Conroy+2010ApJ...712..833C}, we updated the age estimates for 199 \citetalias{Gupta+2011ApJ...740...92G} hosts and 102 \citetalias{Rose+2019ApJ...874...32R} hosts.
Because the sample includes star-forming galaxies, we incorporated the FSPS nebular-emission module and the CLOUDY-based dust prescription \citep{Byler+2017ApJ...840...44B}, which account for non-negligible contributions from nebular emission and dust attenuation to the observed spectral energy distributions (SEDs).

To infer stellar population properties, \PaperI adopted the broadband SED fitting with a Markov Chain Monte Carlo (MCMC)-based methodology following \citetalias{Rose+2019ApJ...874...32R}, in which parameterized star-formation histories (SFHs) are used.
Broadband SED-based age inference remains subject to substantial uncertainty because of the well-known degeneracies among stellar age, metallicity, dust attenuation, and assumed SFH \citep[e.g.,][]{Worthey1994ApJS...95..107W, O_Connell1996ASPC...98....3O, Bruzual&Charlot2003MNRAS.344.1000B, Pforr+2012MNRAS.422.3285P, Conroy2013ARA&A..51..393C, Pacifici+2016ApJ...832...79P}.
Although the MCMC-based approach does not eliminate these intrinsic degeneracies, it provides a systematic framework to propagate them into posterior constraints on the inferred host properties and their uncertainties.
Within this framework, \PaperI showed that different SFH assumptions produce only modest changes in the median host ages, indicating that these estimates remain suitable for relative comparisons among host populations.
We therefore derived host ages and metallicities from the posterior SFHs that reproduce the observed photometric SEDs.
These host ages should be regarded as model-dependent estimates of mass-weighted host stellar population age rather than direct measurements of the SN Ia progenitor age.



\subsection{Strong Progenitor Age Dependence in the Phillips Relation}
\label{sec2.2.phillips}

\begin{figure*}
    \centering
    \includegraphics[width=0.9\linewidth]{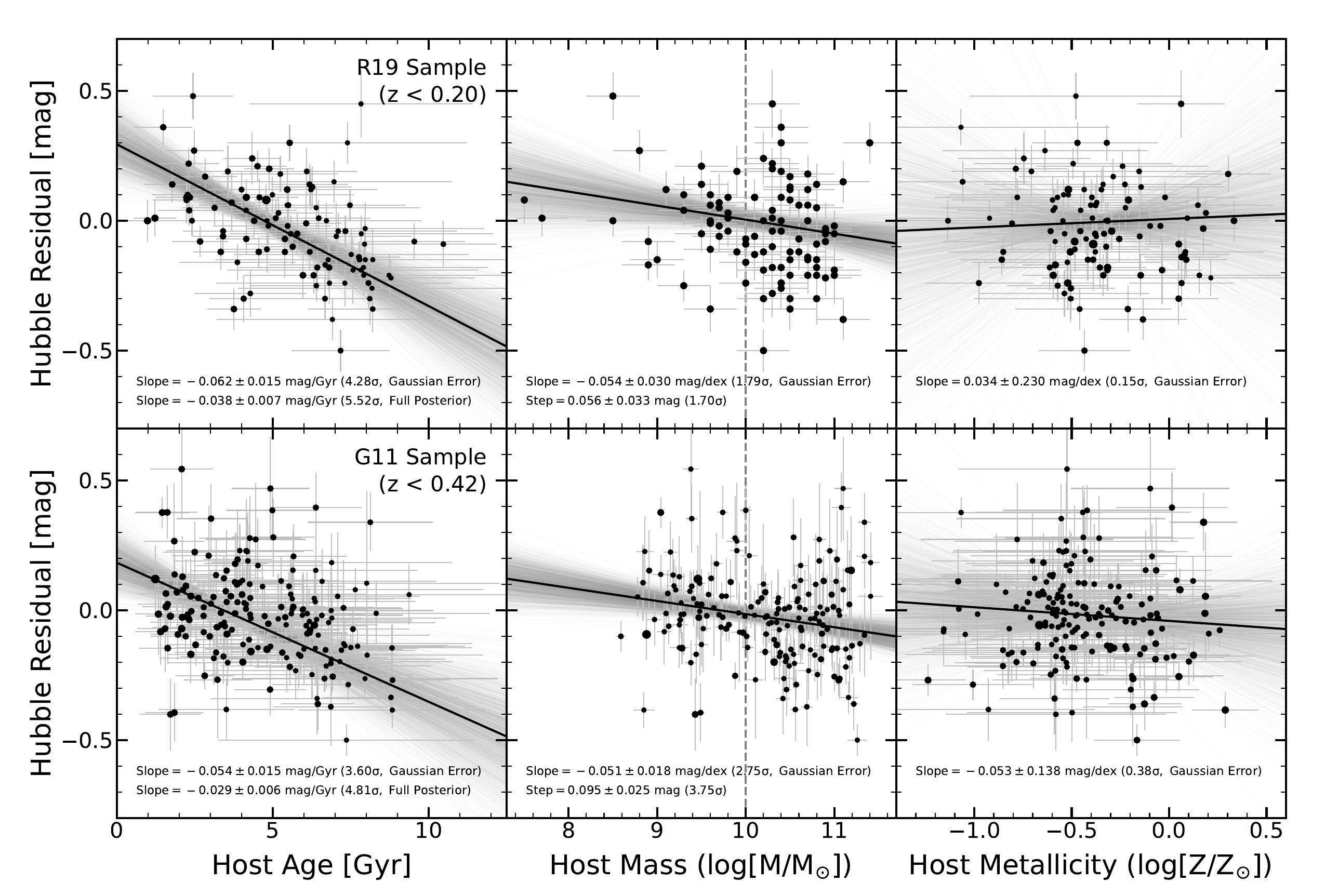}
    \caption
    {Correlations between the HR and three host properties---age, mass, and metallicity---for the \citetalias{Rose+2019ApJ...874...32R} (top panels) and \citetalias{Gupta+2011ApJ...740...92G} (bottom panels) samples.
    The HR is used as a measure of relative SN Ia luminosity after standardization (see text for the redshift-evolution correction applied to the \citetalias{Gupta+2011ApJ...740...92G} sample).
    Among the host properties considered, the age--HR correlation is clearly the strongest.
    The gray lines show Markov chain Monte Carlo (MCMC) linear regressions obtained using the LinMix package \citep{Kelly2007ApJ...665.1489K}, while the thick black lines indicate the corresponding best-fit relations.
    The fitted slopes and their statistical significances derived using LinMix are listed as `Gaussian Error' at the bottom of each panel.
    For the age–HR relation, results derived from the full age posterior distributions are adopted (denoted by `Full Posterior').
    For the host-mass panels, the amplitudes of the mass-step are also indicated.
    Data-point sizes in each panel are weighted by the inverse uncertainties in host age, mass, and metallicity, respectively.}
    \label{fig1}
\end{figure*}

\begin{figure*}
    \centering
    \includegraphics[width=0.7\linewidth]{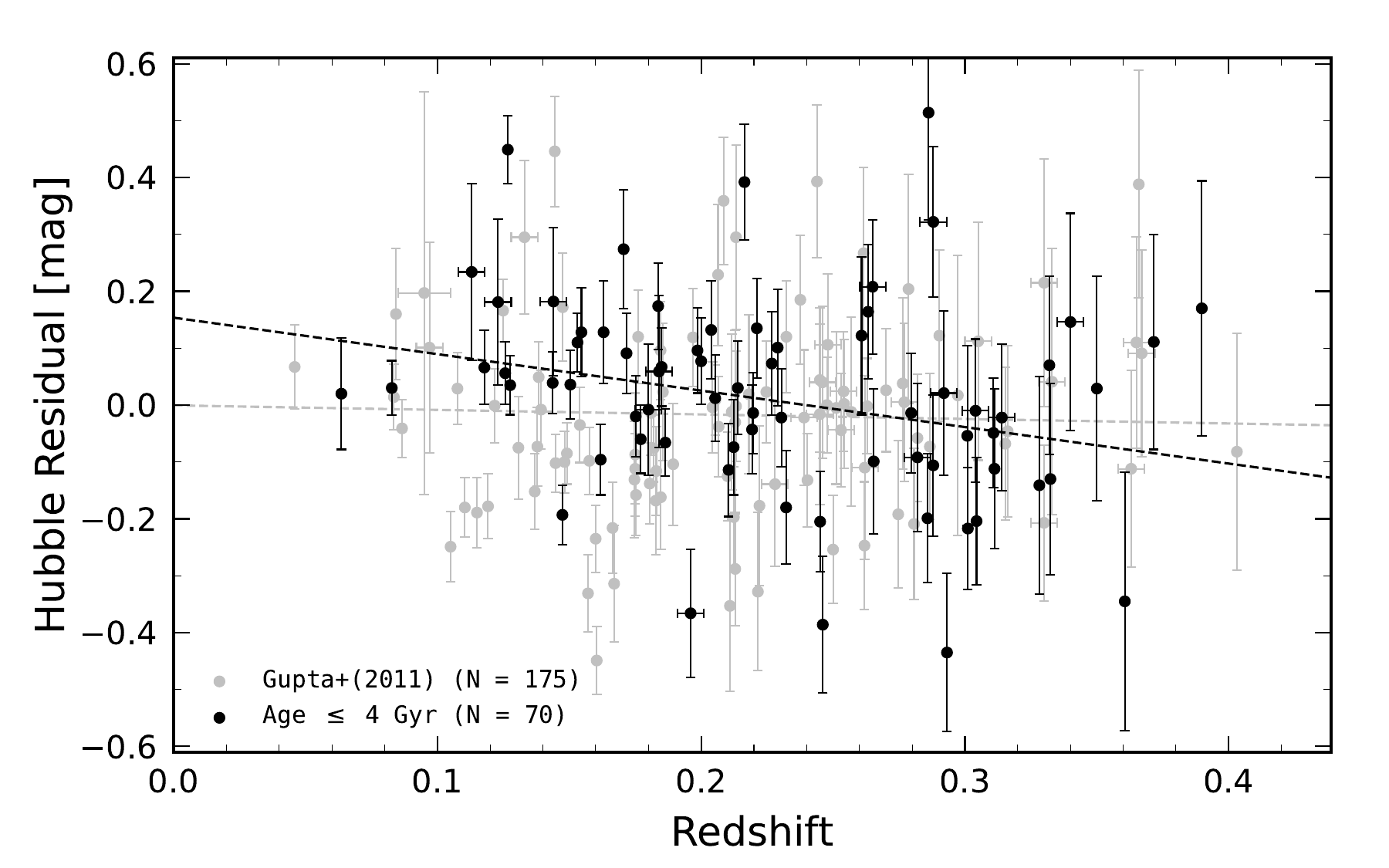}
    \caption
    {Redshift evolution of HR within the \citetalias{Gupta+2011ApJ...740...92G} sample ($z < 0.42$).
    The HRs of 70 SNe Ia originating from young and approximately coeval host galaxies (age $\leq$ 4~Gyr; black dots), selected from the 175 SNe Ia in the \citetalias{Gupta+2011ApJ...740...92G} sample (gray dots), exhibit a clear redshift-dependent trend.
    If this trend is not corrected for, the inferred slope of the host-age--HR regression is severely underestimated (see text).}
    \label{fig2}
\end{figure*}

\begin{figure*}
    \centering
    \includegraphics[width=1.0\linewidth]{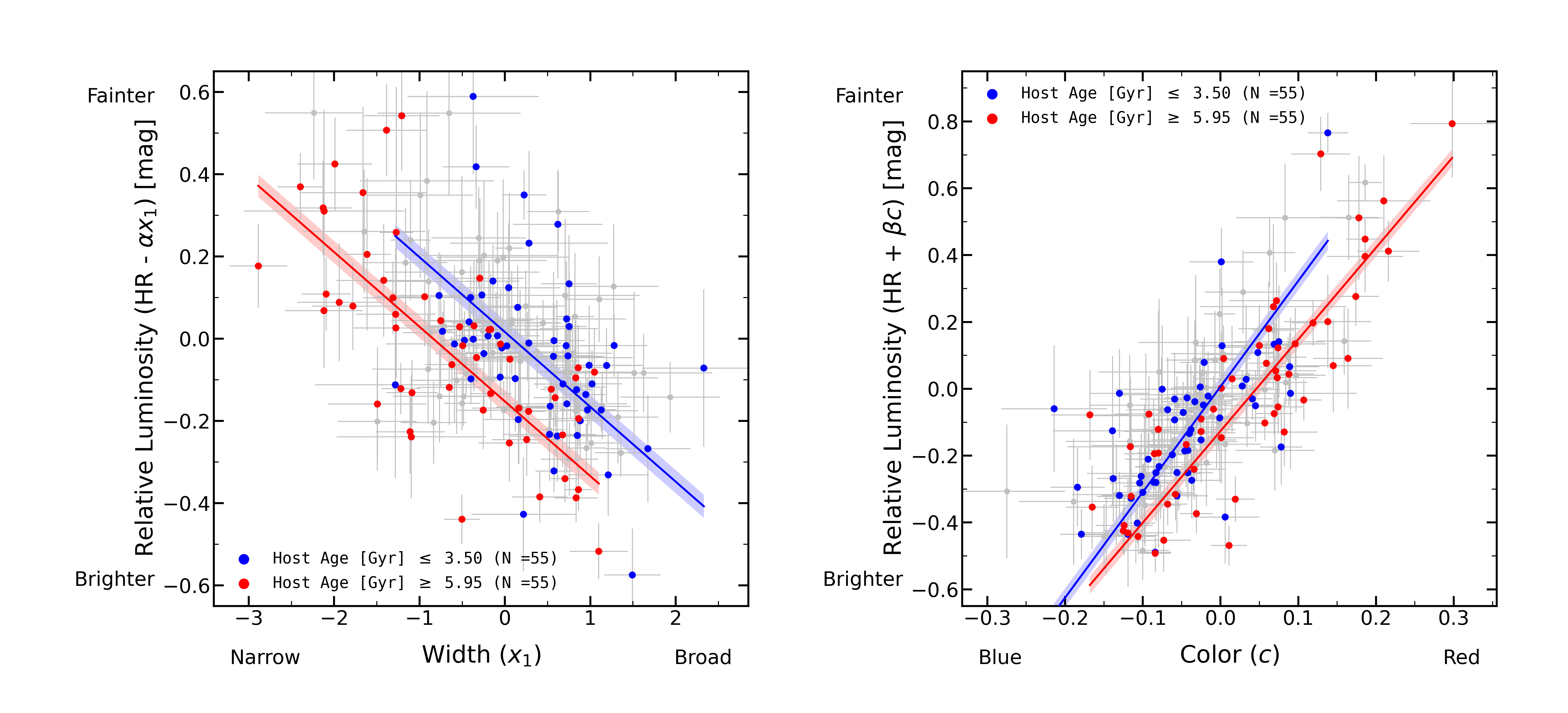}
    \caption
    {Strong host-age dependence of the Phillips relation (width--luminosity relation; WLR; left) and the color--luminosity relation (CLR; right) for SNe Ia in the \citetalias{Gupta+2011ApJ...740...92G} sample.
    The sample is divided into young (blue circles) and old (red circles) subgroups, with an intermediate gray zone.
    Solid lines show best-fit regressions, and shaded bands indicate the $1\sigma$ uncertainty in the intercept from MCMC posterior sampling \citep{Kelly2007ApJ...665.1489K}.
    SNe Ia from younger progenitors are fainter at a given width ($x_1$) and color ($c$), consistent with the result from the \citetalias{Rose+2019ApJ...874...32R} sample (\citetalias{Lee+2022MNRAS.517.2697L}).}
    \label{fig3}
\end{figure*}

\Cref{fig1} shows the correlations between the Hubble residual (HR)\footnote{The HR is defined as the difference between the observed ($\mu_\text{SN}$) and model-predicted ($\mu_\text{model}$) distance moduli ($\text{HR} \equiv \mu_\text{SN} - \mu_\text{model}$).} and three host properties---age, mass, and metallicity---where HR is used as a measure of relative SN Ia luminosity after standardization.
Our primary goal here is to examine how HR varies as a function of host properties.
For this purpose, the HR values must not include any additional corrections based on host properties, such as the host mass-step correction, beyond the standard light-curve width ($x_1$) and color ($c$) corrections.
We therefore adopt the HR values as originally determined by the authors of the \citetalias{Rose+2019ApJ...874...32R} and \citetalias{Gupta+2011ApJ...740...92G} samples.
Host-galaxy stellar-mass measurements are also taken from \citetalias{Rose+2019ApJ...874...32R} and \citetalias{Gupta+2011ApJ...740...92G}.
As shown in \Cref{fig1}, among the host properties considered, the age–HR correlation is clearly the strongest.
The \citetalias{Gupta+2011ApJ...740...92G} sample lies at relatively higher redshift, where larger photometric uncertainties lead to increased errors in both the measured ages and HR values, resulting in a slightly lower statistical significance compared to the \citetalias{Rose+2019ApJ...874...32R} sample.
Nevertheless, the overall trend is fully consistent with that observed in the \citetalias{Rose+2019ApJ...874...32R} sample.
When the full age posterior distributions---derived from the MCMC-based FSPS age estimation and thus naturally propagating uncertainties in the stellar population models---are incorporated into the regression analysis, the statistical significances of the age–HR correlations increase to $5.52\sigma$ and $4.81\sigma$ for the \citetalias{Rose+2019ApJ...874...32R} and \citetalias{Gupta+2011ApJ...740...92G} samples, respectively.

For the \citetalias{Gupta+2011ApJ...740...92G} sample, this analysis differs from \PaperI only in that, following standard practice, we apply quality cuts to the light-curve fit parameters. 
The selection criteria are $|x_1| \leq 3$ and $|\sigma_{x_1}| \leq 1$, and $|c| \leq 0.3$ and $|\sigma_c| \leq 0.1$.
After applying these data selection filters, the final sample consists of 175 objects out of the original 199.
Unlike the \citetalias{Rose+2019ApJ...874...32R} sample, which is confined to a narrow redshift range ($0.04 < z < 0.20$), the \citetalias{Gupta+2011ApJ...740...92G} sample spans a much wider interval ($0.08 < z < 0.42$), such that the HR cannot be interpreted purely as a measure of relative luminosity, but instead acquires a redshift-dependent contribution from the assumed baseline cosmological model.
Therefore, it is necessary to correct for the redshift evolution of HR within this interval.
To do so, we follow the method described in \PaperI and examine the redshift dependence using only a subsample with similar host galaxy ages.
\Cref{fig2} shows that the HRs of 70 SNe Ia originating from young host galaxies (age $\leq$ 4~Gyr), selected from the 175 SNe Ia in the filtered \citetalias{Gupta+2011ApJ...740...92G} sample, exhibit a clear redshift-dependent trend.
A linear regression fit to this young subsample (black dashed line) was performed using the LinMix package \citep{Kelly2007ApJ...665.1489K}, yielding $\text{HR} = (-0.641 \pm 0.284) \, z + (0.154 \pm 0.063) \, \mathrm{[mag]}$.
However, a fit to the entire sample (gray dashed line) yields a negligible slope, $\text{HR} = (-0.080 \pm 0.189) \, z + (0.001 \pm 0.041) \, \mathrm{[mag]}$.
To correct for this redshift evolution, we apply a redshift-dependent correction to the HRs of the full \citetalias{Gupta+2011ApJ...740...92G} sample, using the offsets between the best-fit regression for the young subsample and the zero-point (HR = 0).
In this context, we take the opportunity to emphasize that several studies investigating the correlation between host age and HR, using samples that span a relatively wide redshift range, have not applied such redshift-evolution corrections, which can lead to a systematic underestimation of the inferred slope.
For example, very recently, \citet{Wiseman+2026arXiv260113785W} claimed that the slope of the host-age--HR correlation in a combined \citetalias{Rose+2019ApJ...874...32R} + \citetalias{Gupta+2011ApJ...740...92G} sample is substantially weaker, and of low statistical significance, compared to that derived in our \Cref{fig1}.
However, the redshift interval of their combined sample ($0.04 < z < 0.42$) is even broader than that of the \citetalias{Gupta+2011ApJ...740...92G} sample alone, and over this range, the redshift evolution of the mean host age reaches $\sim$\,3.5~Gyr \citep[see Paper II in this series;][hereafter \citetalias{Son+2025MNRAS.544..975S}]{Son+2025MNRAS.544..975S}.
Therefore, SNe Ia spanning host-age differences of up to $\sim$\,3.5~Gyr ($\sim$~40\% of the full age span) were effectively assigned the same HR values even before performing the host-age--HR regression.
This procedure inevitably leads to a severely underestimated slope.

To investigate more thoroughly the origin of the strong correlation with host age, and why this correlation is much stronger than those with other host properties, we follow \citet[hereafter \citetalias{Lee+2022MNRAS.517.2697L}]{Lee+2022MNRAS.517.2697L} and examine the effects of host age, mass, and metallicity on the width--luminosity relation (WLR, the ``Phillips relation'') and the color--luminosity relation (CLR) in the SN Ia luminosity standardization process \citep{Phillips1993ApJ...413L.105P, Tripp1998A&A...331..815T}.
These two correction terms are incorporated into the calculation of the SN Ia distance modulus, 
\begin{equation}
    \mu_{\text{SN}} = m - M + \alpha x_1 - \beta c \, ,
    \label{eq1}
\end{equation}
where $m$ and $M$ are the apparent and absolute magnitudes, $x_1$ and $c$ are the light-curve width and color parameters as defined in the SALT2 model \citep{Guy+2007A&A...466...11G}, and $\alpha$ and $\beta$ are the absolute values of the slopes of the WLR and CLR, respectively.
Using the \citetalias{Rose+2019ApJ...874...32R} sample, \citetalias{Lee+2022MNRAS.517.2697L} demonstrated that the WLR and CLR exhibit a strong dependence on progenitor age, with the associated magnitude offsets detected at a significance level of $4.61\sigma$.
This result is contrary to the key assumption of SN cosmology \citep{Jha+2019NatAs...3..706J} that these relations should be invariant with progenitor age.
In contrast, when the sample is divided according to other host properties, such as stellar mass, metallicity, or dust content, the resulting offsets in the WLR and CLR are statistically insignificant ($0.98\sigma$--$1.25\sigma$), indicating that these properties are unlikely to be the primary drivers of the observed correlations between host properties and HR.
Using the newly measured ages presented in \PaperI, we confirm that these results remain essentially unchanged.
In this paper, we further extend the same analysis to the \citetalias{Gupta+2011ApJ...740...92G} sample at relatively higher redshift to further test the robustness of these conclusions.

\Cref{fig3} presents the WLR and CLR of SNe Ia in the \citetalias{Gupta+2011ApJ...740...92G} sample using our newly measured host ages from \PaperI.
Following \citet{Astier+2006A&A...447...31A}, the HRs in the WLR are computed using the observed distance $\mu_\text{SN}$ without applying the light-curve width correction term $\alpha x_1$, i.e., corrected only for color $c$.
Conversely, the HRs in the CLR are computed without the color correction term $\beta c$, applying only the width correction term $\alpha x_1$.
We adopt the standardization coefficients $(\alpha, \beta) = (0.121, 2.82)$, as fitted by \citetalias{Gupta+2011ApJ...740...92G}.
To investigate the host age dependence in the WLR and CLR,  we divide the sample into young and old subgroups, introducing a gray zone in between to minimize cross-contamination.
The age ranges are defined such that the young subsample, the gray zone, and the old subsample each contain approximately one third ($\sim 33.3\%$) of the total sample.
Based on this division, we repeat the same experiment while varying the number of SNe Ia in each subgroup by $\pm 3$ percentage points and then derive averaged results.
With this scheme, the width of the gray zone becomes comparable to the typical age measurement uncertainty of $\sim$\,1.9~Gyr.

\Cref{fig3} shows one representative example illustrating the approximate mean outcome from these repeated experiments.
As in the \citetalias{Rose+2019ApJ...874...32R} sample analyzed by \citetalias{Lee+2022MNRAS.517.2697L}, a clear host age dependence is also observed in the WLR and CLR for the \citetalias{Gupta+2011ApJ...740...92G} sample.
The differences in the slopes of the calibrating relations between the two age subgroups are statistically insignificant, with $(\Delta\alpha, \Delta\beta) = (0.000 \pm 0.054, 0.418 \pm 0.494)$.
However, at $x_1 = 0$, the zero-point of the WLR for the young subgroup is shifted toward fainter magnitudes by $0.169 \pm 0.039 \, \text{mag} \ (4.33\sigma)$ relative to that of the old subgroup.
The average zero-point shift measured across all subdivision trials is $0.153 \pm 0.038 \, \text{mag}$, corresponding to an average statistical significance of $4.04\sigma$ level.
The variance of the zero-point shifts obtained from different subgroup combinations is an order of magnitude smaller than their squared measurement uncertainties.
This indicates that additional scatter arising from the choice of subdivision is negligible, and that the inferred bias is statistically consistent across all tested subdivision criteria.
As emphasized by \citetalias{Lee+2022MNRAS.517.2697L}, this zero-point offset of SN Ia luminosity is reminiscent of \citet{Baade1956PASP...68....5B}'s discovery of two Cepheid period--luminosity relations that vary with stellar population age---a finding that revealed the Hubble constant originally determined by \citet{Hubble1929PNAS...15..168H} to have been substantially misestimated---and would therefore introduce a serious systematic bias in SN cosmology.
Specifically, SNe Ia originating from younger progenitors are systematically overcorrected by the standardization procedure and thus appear fainter after calibration (see left panel of \Cref{fig5}).
Importantly, SNe Ia observed at higher redshift predominantly arise from younger progenitor populations than those at low redshift.
As a result, part of the observed dimming of high-redshift SNe Ia relative to a baseline model without a cosmological constant can be attributed to this age-dependent overcorrection in the luminosity standardization process (see \PaperII).

As shown in \Cref{fig3}, SNe Ia from younger progenitors appear fainter at a given light-curve width and color, yet the younger subgroup is overall brighter when the age-defined subsamples are compared directly.
This result is supported by current explosion and radiative-transfer models of SNe Ia \citep[e.g.,][]{Lesaffre+2006MNRAS.368..187L, Krueger+2010ApJ...719L...5K, Krueger+2012ApJ...757..175K, Seitenzahl+2013MNRAS.429.1156S}.
Although theoretical models for SNe Ia are still incomplete, it is well established that their peak luminosities increase with the amount of radioactive $^{56}$Ni synthesized in the explosion, which itself tends to increase with progenitor mass \citep[e.g.,][]{Woosley+2007ApJ...662..487W, Leung+2021ApJ...909..152L}.
Therefore, the differences observed in \Cref{fig3} can be understood if younger progenitors correspond, on average, to higher-mass systems that synthesize more $^{56}$Ni, and hence give rise to intrinsically more luminous SNe Ia with broader and bluer light curves (see also \citetalias{Lee+2022MNRAS.517.2697L}).
Consistent with this interpretation, SNe Ia in younger hosts are associated with higher $^{56}$Ni production, supporting a more direct role for progenitor age \citep{Howell+2009ApJ...691..661H}.
Taken together, these considerations suggest that progenitor age is a plausible physical driver of the observed luminosity variations, although a quantitative assessment will require more detailed modeling.

\begin{figure*}
    \centering
    \includegraphics[width=0.9\linewidth]{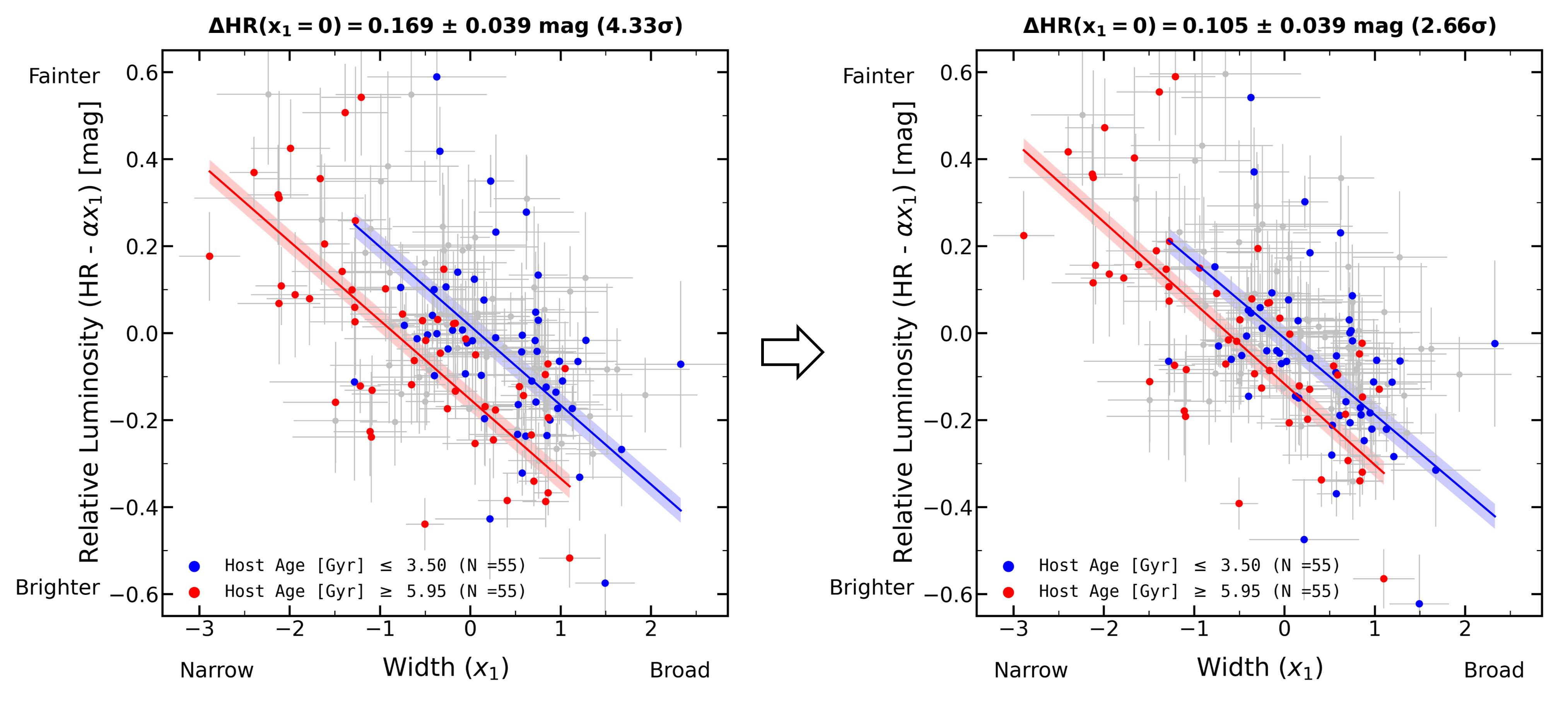}
    \caption
    {Same as \Cref{fig3}, but comparing the Phillips relation before (left) and after (right) applying the host mass-step correction.
    Even after the mass-step correction, a statistically meaningful offset remains between the young and old subgroups.}
    \label{fig4}
\end{figure*}

\begin{figure*}
    \centering
    \includegraphics[width=0.9\linewidth]{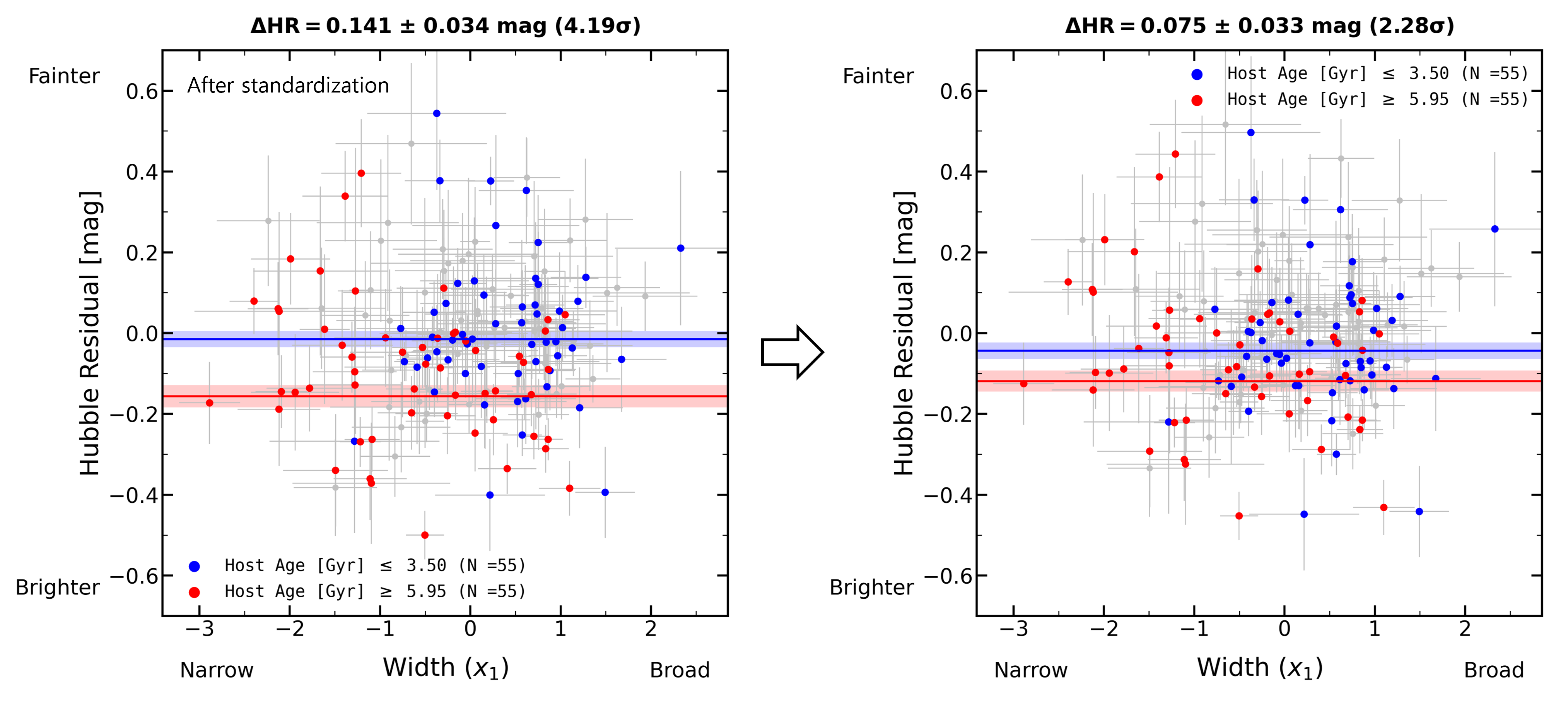}
    \caption
    {Same as \Cref{fig4}, but after applying the full SN Ia luminosity standardization.
    This shows that the host mass-step correction only partially suppresses the progenitor-age dependence in standardized SN Ia magnitudes.}
    \label{fig5}
\end{figure*}

\begin{figure*}
    \centering
    \includegraphics[width=0.9\linewidth]{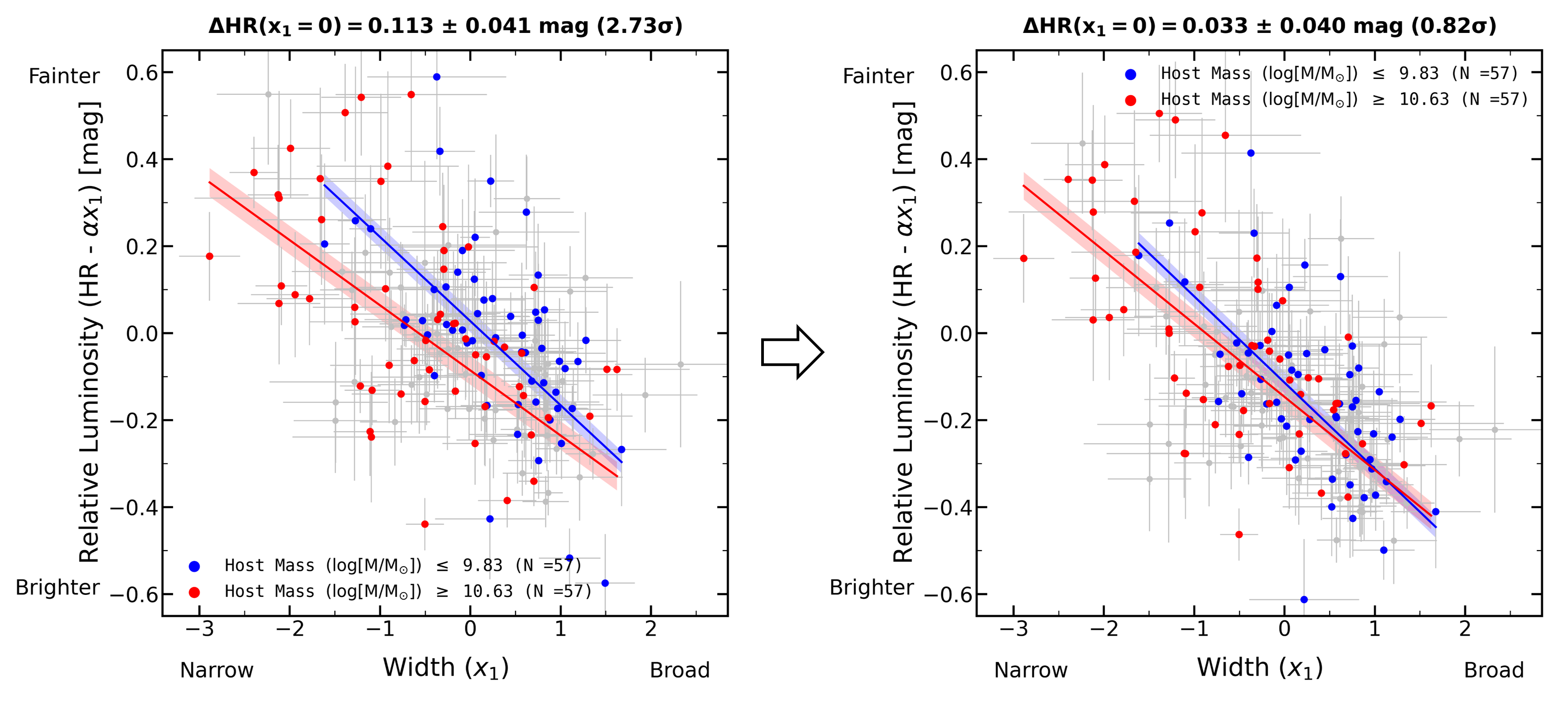}
    \caption
    {Host-mass dependence of the Phillips relation, shown before (left) and after (right) the age-bias correction.
    Even before the age-bias correction, the luminosity offset between the low- and high-mass subgroups is smaller than the offset between the young and old subsamples in \Cref{fig3}.
    After applying the age-bias correction, the two mass-defined subgroups converge and the luminosity offset effectively vanishes.}
    \label{fig6}
\end{figure*}

\begin{figure*}
    \centering
    \includegraphics[width=0.9\linewidth]{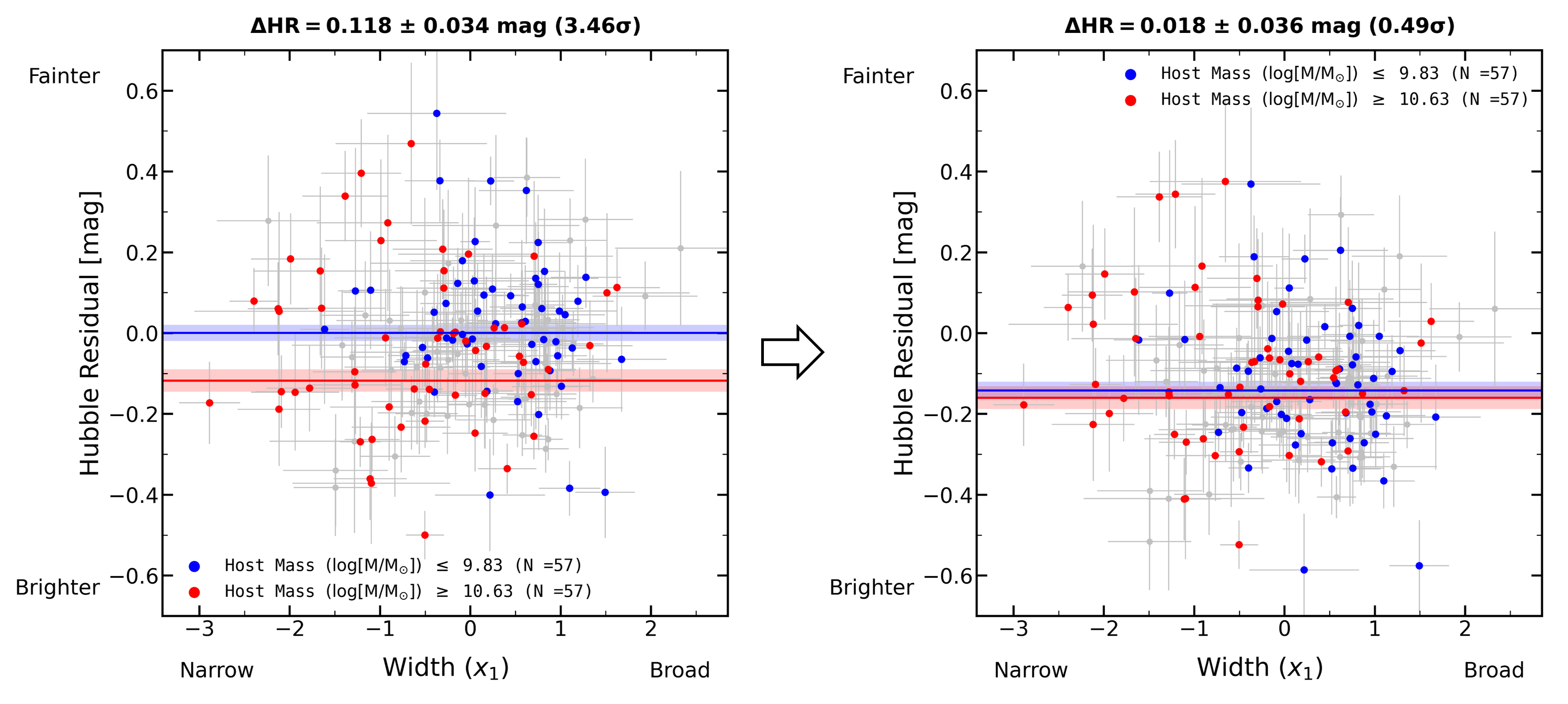}
    \caption
    {Same as \Cref{fig6}, but after applying the full SN Ia luminosity standardization.
    This shows that the age-bias correction fully suppresses the host mass step in standardized SN Ia magnitudes.}
    \label{fig7}
\end{figure*}

\begin{figure*}
    \hspace*{\fill} 
    \begin{subfigure}{0.4\linewidth}
        \includegraphics[width=\linewidth]{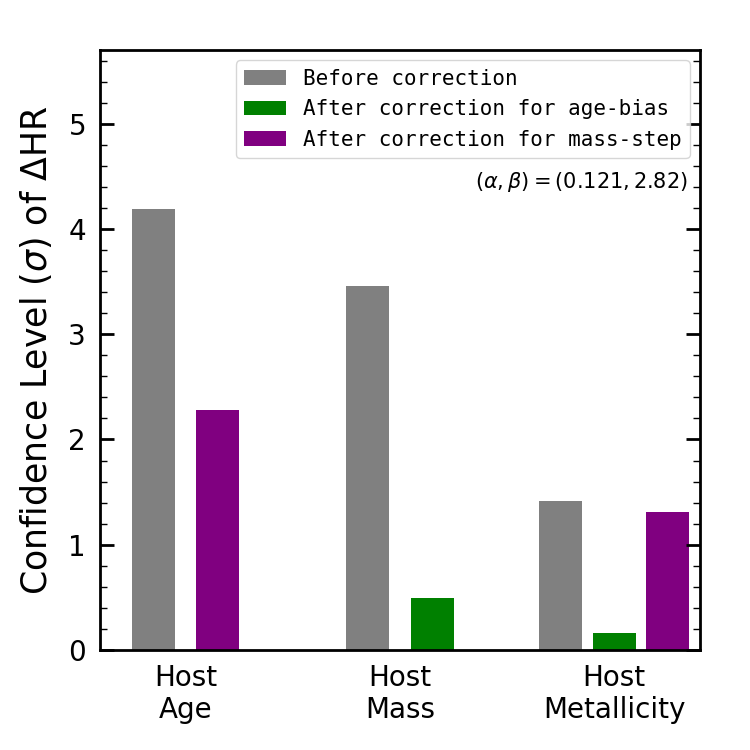}
    \end{subfigure}
    \hfill 
    \begin{subfigure}{0.4\linewidth}
        \includegraphics[width=\linewidth]{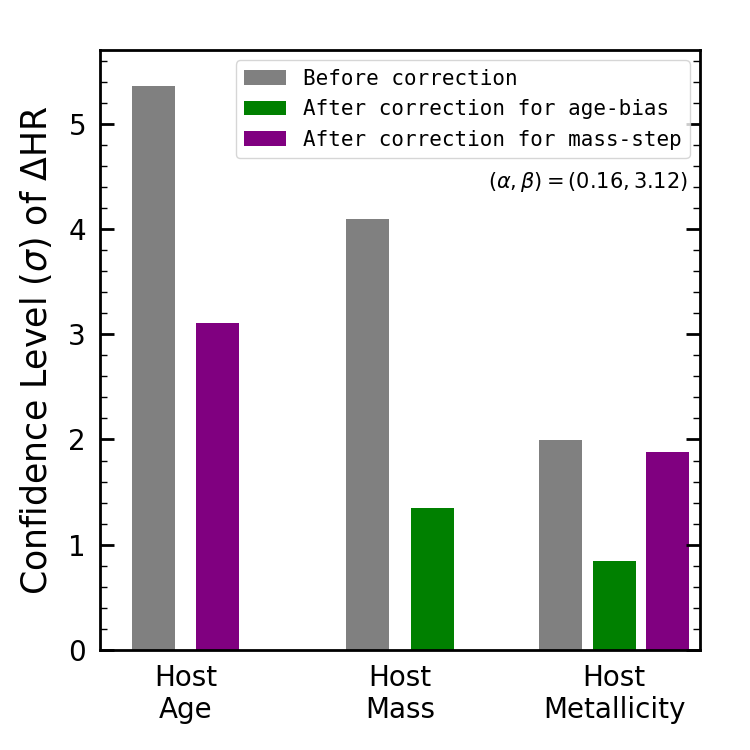}
    \end{subfigure} 
    \hspace*{\fill} 
    \caption
    {\textbf{\textit{Left:}} Confidence levels of the $\Delta \text{HR}$ between two SN Ia subgroups defined by host properties (age, mass, and metallicity).
    Gray bars show results before any correction is applied.
    Purple bars show results after the host mass-step correction, and green bars show results after the age-bias correction.
    The $\Delta \text{HR}$ between the age-defined subgroups is the most significant, exceeding the low- \& high-mass difference both before any correction and after applying either the mass-step or the age-bias correction.
    The $\Delta \text{HR}$ associated with metallicity is not significant ($1.42\sigma$) and is further reduced by the age-bias correction, while the mass-step correction has negligible impact.
    \textbf{\textit{Right:}} Same as left panel, but adopting recently favored values for the SN Ia standardization coefficients, $(\alpha, \beta) = (0.16, 3.12)$.
    The overall changes in confidence levels are similar to left panel, but the confidence levels are generally increased.}
    \label{fig8}
\end{figure*}

Spatial variations in age, metallicity, and star formation activity within galaxies suggest that integrated host-galaxy properties may not fully represent the local environments at SN Ia explosion sites \citep[e.g.,][]{Sanchez+2012A&A...538A...8S, Hakobyan+2012A&A...544A..81H, Hakobyan+2014MNRAS.444.2428H, Bundy+2015ApJ...798....7B, Gonzalez_Delgado+2015A&A...581A.103G}.
In practice, however, we found no noticeable difference in either the slope or the significance of the age-HR correlation between global and local age measurements \citep{Lee+2020ApJ...903...22L, Chung+2025MNRAS.538.3340C}.
Also, the offsets between the two age subgroups in the WLR are very similar whether one uses global or local ages.
This lack of distinction is likely due to the higher uncertainties in local measurements from SDSS, which potentially obscure any subtle environmental differences that might otherwise distinguish them from global properties.

Similarly to the age analysis, we examine whether other host-galaxy properties, host mass and metallicity, exhibit comparable trends in the WLR and CLR.
For consistency, the subgroups defined by host mass and metallicity are constructed to contain the same number of SNe Ia as the age-based subgroups.
In contrast to host age, metallicity shows a statistically insignificant luminosity offset at the $\leq 1\sigma$ level.
Host mass exhibits a larger offset at the $\sim 2.73\sigma$ level, but this significance is still much lower than that associated with host age.
These results are qualitatively consistent with those previously reported by \citetalias{Lee+2022MNRAS.517.2697L} for the \citetalias{Rose+2019ApJ...874...32R} sample.
This suggests that the host mass step is not an independent effect, but rather a secondary manifestation of the underlying progenitor age bias that drives the observed host-property dependence in standardized SN Ia magnitudes.
To further test this interpretation, and to determine which dependence reflects the fundamental origin rather than a derivative effect, we investigate below how the differences between the two subgroups change after correcting the age-defined subgroups for mass effects, and conversely, correcting the mass-defined subgroups for age effects---specifically examining whether the offsets are fully suppressed or remain only modestly affected.



\subsection[Disentangling Progenitor Age Effect from Apparent Mass Step]{Disentangling Progenitor Age Bias from \\ Apparent Host Mass Step}
\label{sec2.2.age&mass}

We first examine how much the age bias is affected when a host mass-step correction is applied in advance.
To this end, following the standard practice in current SN cosmology analyzes \citep[e.g.,][]{DES2024ApJ...973L..14D}, we incorporate a host mass-step correction as an additional term in the SN Ia luminosity standardization process, given by 
\begin{equation}
\begin{split}
    & \mu_{\text{SN,cor}} = \mu_{\text{SN}} + \gamma_M / 2 \, , \\
    & \text{where} \ \gamma_M / 2 = 
    \begin{cases}
        -0.0475\,\text{mag} & \mathrm{for} \ \log\mathrm{[M/M_{\odot}]} < 10 \\
        +0.0475\,\text{mag} & \mathrm{for} \ \log\mathrm{[M/M_{\odot}]} \geq 10 \, .
    \end{cases}
\end{split}
\label{eq2}
\end{equation}
Here we adopt a value of $\gamma_M = 0.095\,\text{mag}$, as derived from the \citetalias{Gupta+2011ApJ...740...92G} sample in \Cref{fig1}.
\Cref{fig4} compares the WLR of \Cref{fig3} before and after the host mass-step correction.
\Cref{fig5} shows the HRs after full SN Ia luminosity standardization is applied to \Cref{fig4}.
Prior to the mass-step correction, SNe Ia from younger host galaxies remain systematically fainter even after luminosity standardization, reflecting the zero-point offset between the two age-defined subgroups seen in the WLR.
To quantify the difference between the two subgroups, we apply Welch’s two-sample \textit{t}-test \citep{Welch1947Biometrika.34.28} with inverse-variance weighting using $1\sigma$ HR uncertainties.
This yields a mean HR difference ($\Delta \text{HR}$) of $0.141 \pm 0.034 \, \text{mag}$, corresponding to a $4.19\sigma$ significance.
The associated p-value ($p = 5.7 \times 10^{-5}$) indicates that the null hypothesis of homogeneity between the young and old subgroups can be firmly rejected, providing strong evidence that their mean HRs are intrinsically different.
After applying the host mass-step correction, the HR difference is reduced to $\Delta \text{HR} = 0.075 \pm 0.033 \, \text{mag}$, but a statistically meaningful offset remains at the $> 97.5\%$ confidence level ($p \approx 0.025$).
This demonstrates that the host mass-step correction only partially suppresses, but does not fully remove, the progenitor age bias in standardized SN Ia magnitudes.
Had we adopted the smaller mass-step amplitude ($\gamma_M = 0.05\,\mathrm{mag}$) employed by \citet{DES2024ApJ...973L..14D}, the reduction in the $\Delta \text{HR}$ would have been correspondingly smaller.

Next, we examine the impact of correcting for the age bias a priori on the host mass step.
To this end, we apply the age–bias slope of $0.029 \, \text{mag/Gyr}$, derived from the \citetalias{Gupta+2011ApJ...740...92G} sample in \Cref{fig1}, to the luminosity standardization process, as described below:
\begin{equation}
\begin{split}
    & \mu_{\mathrm{SN,cor}} = \mu_{\mathrm{SN}} + \Delta m (\mathrm{age}) \, ,\\
    & \mathrm{where} \ \Delta m(\mathrm{age}) = 0.029\,\mathrm{mag}\,\mathrm{Gyr}^{-1} \times (\mathrm{age} - 8.2\,\mathrm{Gyr}) \, .
\end{split}
\label{eq3}
\end{equation}
Following our \PaperII, the relative age difference term in \Cref{eq3} is defined with respect to $z = 0.0$, where the mean host age is 8.2~Gyr; however this choice has no impact on the relative differences in HR.
\Cref{fig6,fig7} present a comparison analogous to \Cref{fig4,fig5}, but with the roles reversed: the sample is subdivided by host mass rather than by host age.
Here, two subsamples are shown for SNe Ia hosted by low-mass ($\log\mathrm{[M/M_{\odot}]} \leq 9.83$) and high-mass ($\log\mathrm{[M/M_{\odot}]} \geq 10.63$) galaxies, using the same subdivision scheme adopted for host age.
In this case, SNe Ia in low-mass hosts appear systematically fainter than those in high-mass hosts, producing a host mass step after standardization.
In \Cref{fig7}, the mean HR difference between the low- and high-mass subgroups is smaller than that between the young and old host subsamples, but remains statistically significant ($\Delta \text{HR} = 0.118 \pm 0.034\,\text{mag}, \ 3.46\sigma$).
After applying the age-bias correction, however, the mean HRs of the two mass-defined subgroups converge, and the HR difference effectively vanishes ($\Delta \text{HR} = 0.018 \pm 0.036\,\text{mag}, \ 0.49\sigma$).
In contrast to the host mass-step correction shown in \Cref{fig5}---which only partially suppresses the age bias---the age-bias correction applied here completely removes the host mass step.
This asymmetry is critical. 
If host mass were the fundamental driver, correcting for mass would comparably suppress the age dependence. 
Instead, the age trend persists, whereas correcting for age removes the mass step, implying that host mass functions only as a secondary proxy.
This strongly indicates that progenitor age is the underlying driver of the observed correlations between host properties and standardized SN Ia magnitudes.

The left panel of \Cref{fig8} summarizes how the $\Delta \text{HR}$ between two subgroups changes after applying corrections for the host mass step and the age bias, when the SN Ia sample is subdivided by host age, mass, and metallicity.
Each bar represents the confidence level of $\Delta \text{HR}$.
Gray bars show $\Delta \text{HR}$ before any correction, while the purple and green bars correspond to values after applying the host mass-step correction and the progenitor age-bias correction, respectively.
The largest and most statistically significant difference is found when the sample is divided by host age ($4.19\sigma$).
Notably, this difference remains substantial ($2.28\sigma$) even after applying the host mass-step correction.
By contrast, the $\Delta \text{HR}$ between the two host mass subsamples is significant at the $3.46\sigma$ level, but becomes negligible ($0.49\sigma$) once the age-bias correction is applied.
The $\Delta \text{HR}$ obtained by subdividing the sample according to host metallicity is already marginal ($1.42\sigma$), and is further suppressed by the age-bias correction, while the host mass-step correction has virtually no effect.
We repeat the same test using a different set of luminosity standardization coefficients, $(\alpha, \beta) = (0.16, 3.12)$, which are commonly adopted in recent SN Ia studies such as the \citet{DES2024ApJ...973L..14D}.
The results are summarized in the right panel of \Cref{fig8}, and the effects of the corrections are fully consistent with those shown in the left panel.
Adopting these more recent coefficients generally increases the statistical significance of $\Delta \text{HR}$.
Taken together, these results strongly indicate that progenitor age is the primary physical driver of the observed host-property dependencies in standardized SN Ia magnitudes.
We adopt this conclusion as the basis for the subsequent analyzes presented in the following section.



\section[Nonlinear Relations between Progenitor Age \& Host Properties as the Origin of SN Ia Magnitude Steps in Hubble Residuals]{Nonlinear Relations between \\ Progenitor Age \& Host Properties \\ as the Origin of SN Ia Magnitude Steps}
\label{sec3.nonlinearity}

The correlations between standardized SN Ia magnitudes and host-galaxy properties (mass and sSFR) are not observed to be simply linear---instead, they exhibit a distinctly nonlinear, step-like behavior. 
To understand this phenomenon, we now present a theoretical demonstration that SN Ia progenitor age provides a unified physical origin for the host-mass and host-sSFR magnitude steps in the HR. 
The purpose of these simulations is not to tune a model to match the observed magnitude steps, but to show that adopting the age--HR slope from \Cref{sec2.root} naturally reproduces the host mass and sSFR steps without additional free parameters. 
In this sense, our simulations provide a predictive consistency check rather than a fit to the observed steps.
Using the framework of \citet[hereafter \citetalias{Childress+2014MNRAS.445.1898C}]{Childress+2014MNRAS.445.1898C},\footnote{We note that binary evolution is an important ingredient in the physics of SN Ia progenitors. However, the present work does not include an explicit treatment of binary population synthesis or progenitor channels \citep[e.g.,][]{Eldridge+2016MNRAS.462.3302E, Eldridge+2017PASA...34...58E}. A more physically complete connection between host-galaxy stellar populations and SN Ia progenitor physics will require such modeling, which we leave to future work.} we model progenitor ages in the form of SN Ia progenitor-age distributions (SPADs, see \citetalias{Childress+2014MNRAS.445.1898C}) and examine their dependence on host-galaxy properties.
We demonstrate that these host property dependencies manifest as a highly nonlinear behavior of progenitor age along the mass and sSFR sequences.
When coupled with a linear relation between progenitor age and HR, the nonlinearity naturally reproduces the observed host-mass and sSFR magnitude steps.



\subsection{Modeling SN Ia Progenitor Ages}
\label{sec3.1.model}

\begin{figure}
    \centering
    \includegraphics[width=\linewidth]{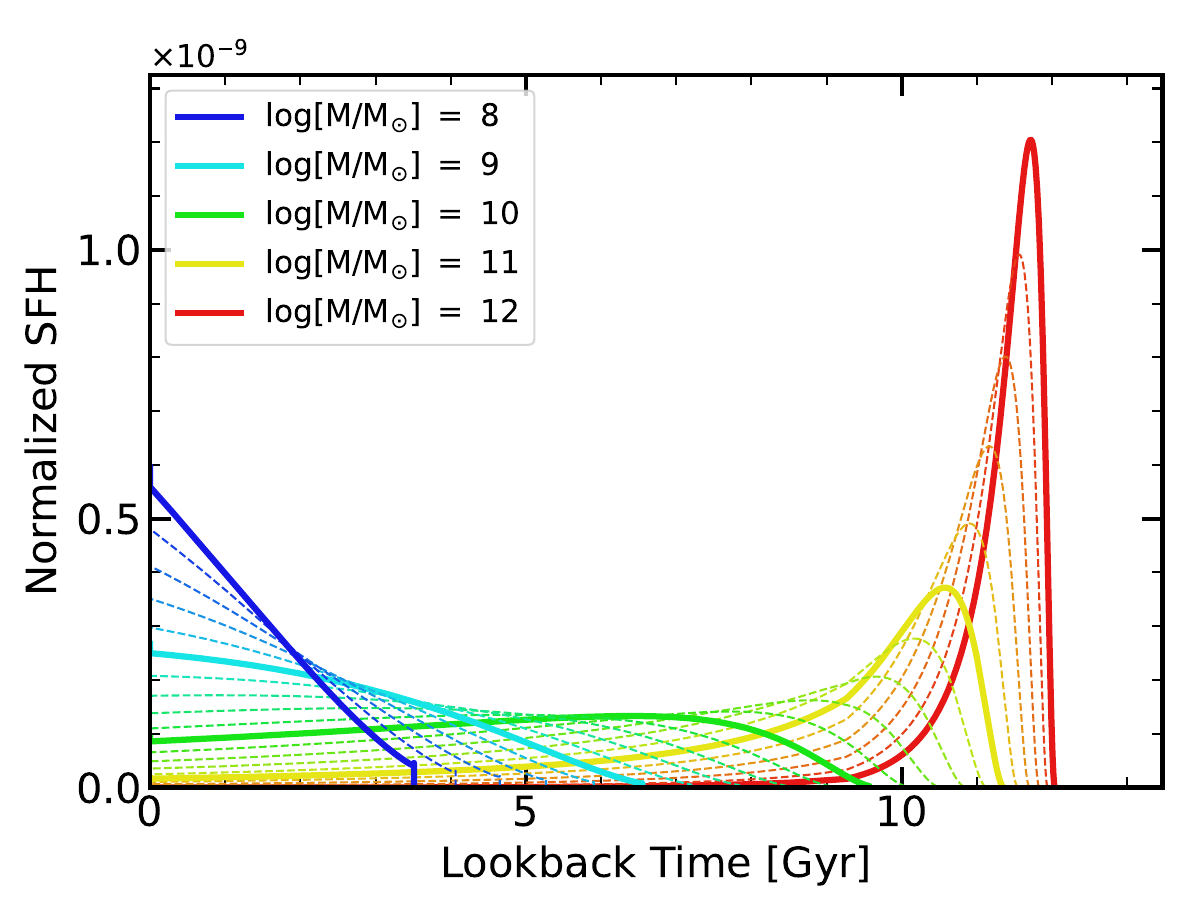}
    \caption
    {SFHs along the galaxy stellar-mass sequence at $z = 0$, computed using the empirical models of \citetalias{Childress+2014MNRAS.445.1898C}.
    star formation rates (SFRs) are normalized by stellar mass (i.e., area-normalized) as functions of look-back time.
    The SFHs are shown as solid and dashed curves for $\log\mathrm{[M/M_{\odot}]} = 8 - 12$ in steps of 0.2~dex.}
    \label{fig9}
\end{figure}

\begin{figure}
    \centering
    \begin{subfigure}{0.5\linewidth}
        \includegraphics[width=\linewidth]{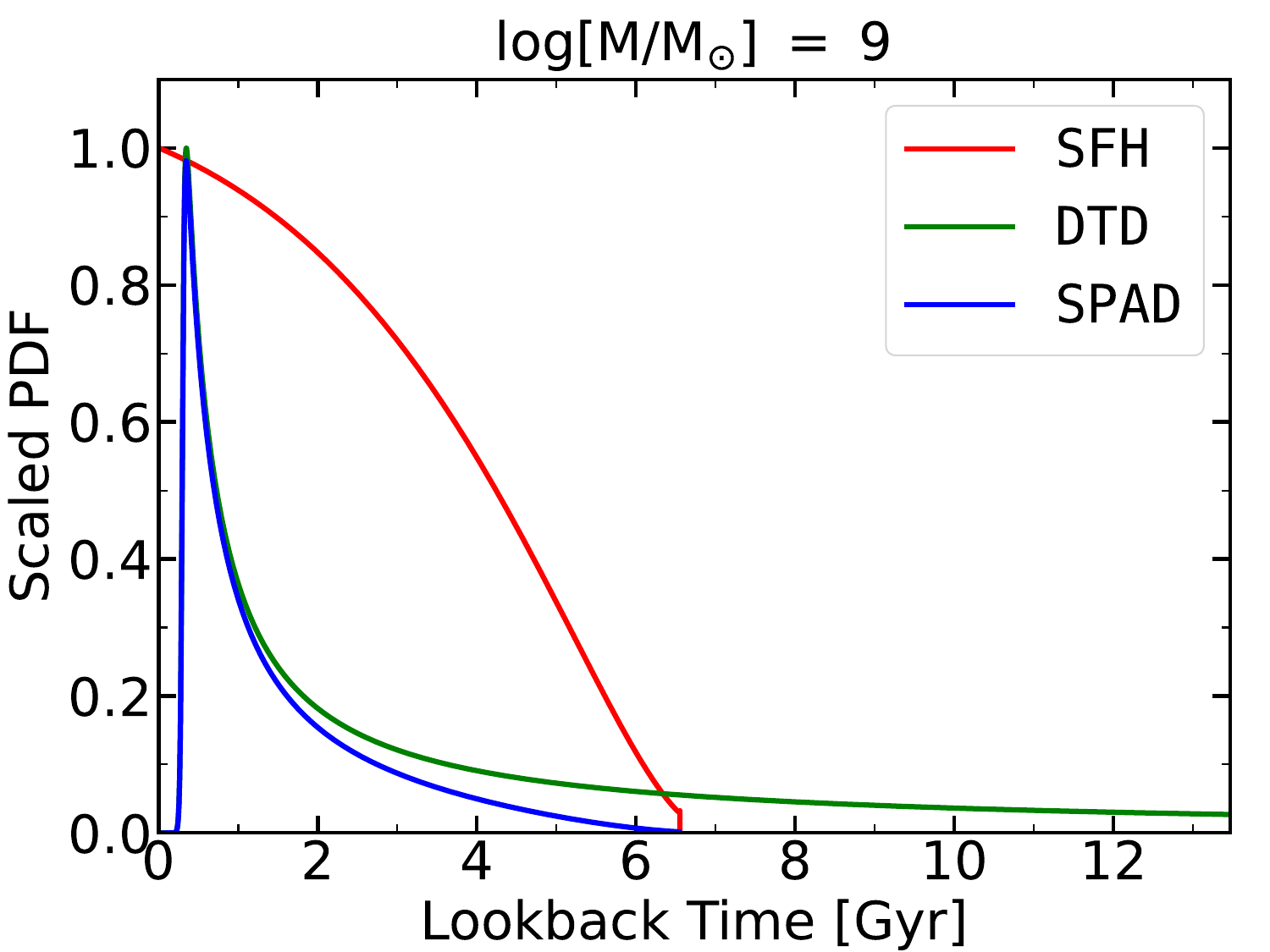}
    \end{subfigure}\hfill 
    \begin{subfigure}{0.5\linewidth}
        \includegraphics[width=\linewidth]{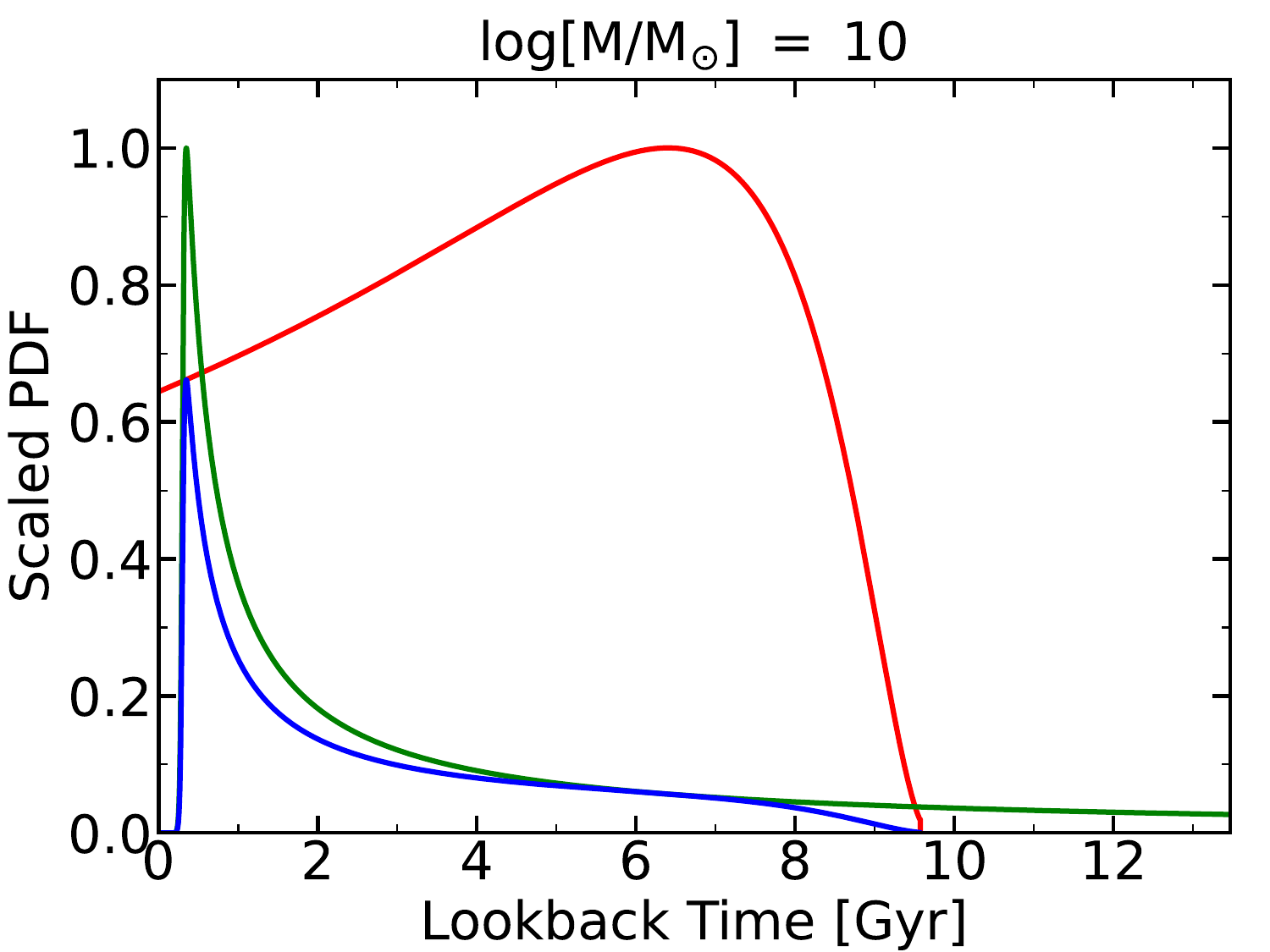}
    \end{subfigure}\\ 
    \begin{subfigure}{0.5\linewidth}
        \includegraphics[width=\linewidth]{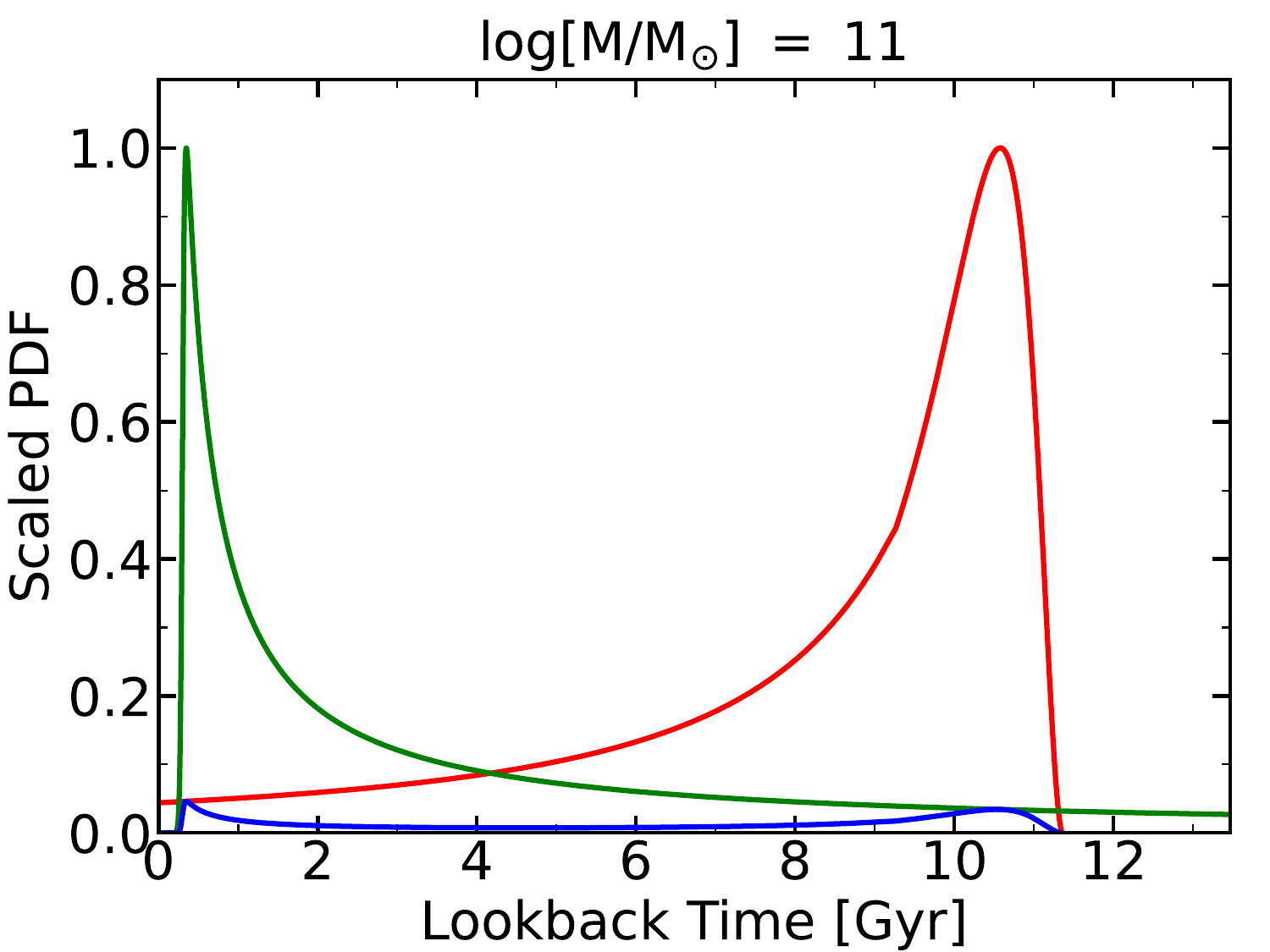}
    \end{subfigure}\hfill 
    \begin{subfigure}{0.5\linewidth}
        \includegraphics[width=\linewidth]{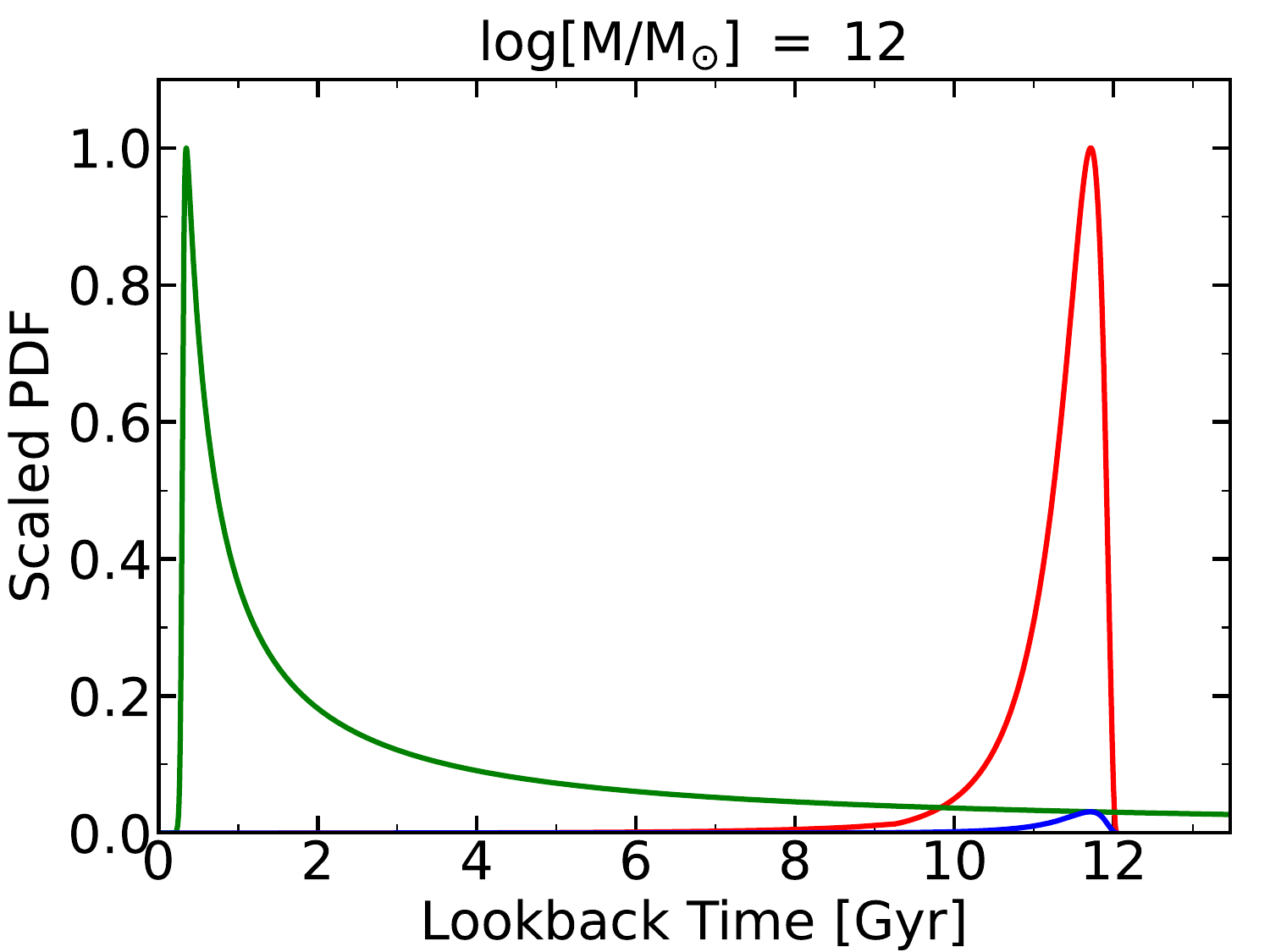}
    \end{subfigure}
    \caption
    {Scaled probability density functions (PDFs) for the galaxy SFHs (red), the DTD (green), and the resulting SPADs (blue). 
    The SFHs are identical to those shown in \Cref{fig9}. 
    The DTD is assumed to be universal and is held fixed across all four panels. 
    The SPADs are computed by convolving each SFH with the DTD.}
    \label{fig10}
\end{figure}

\begin{figure}
    \centering
    \includegraphics[width=\linewidth]{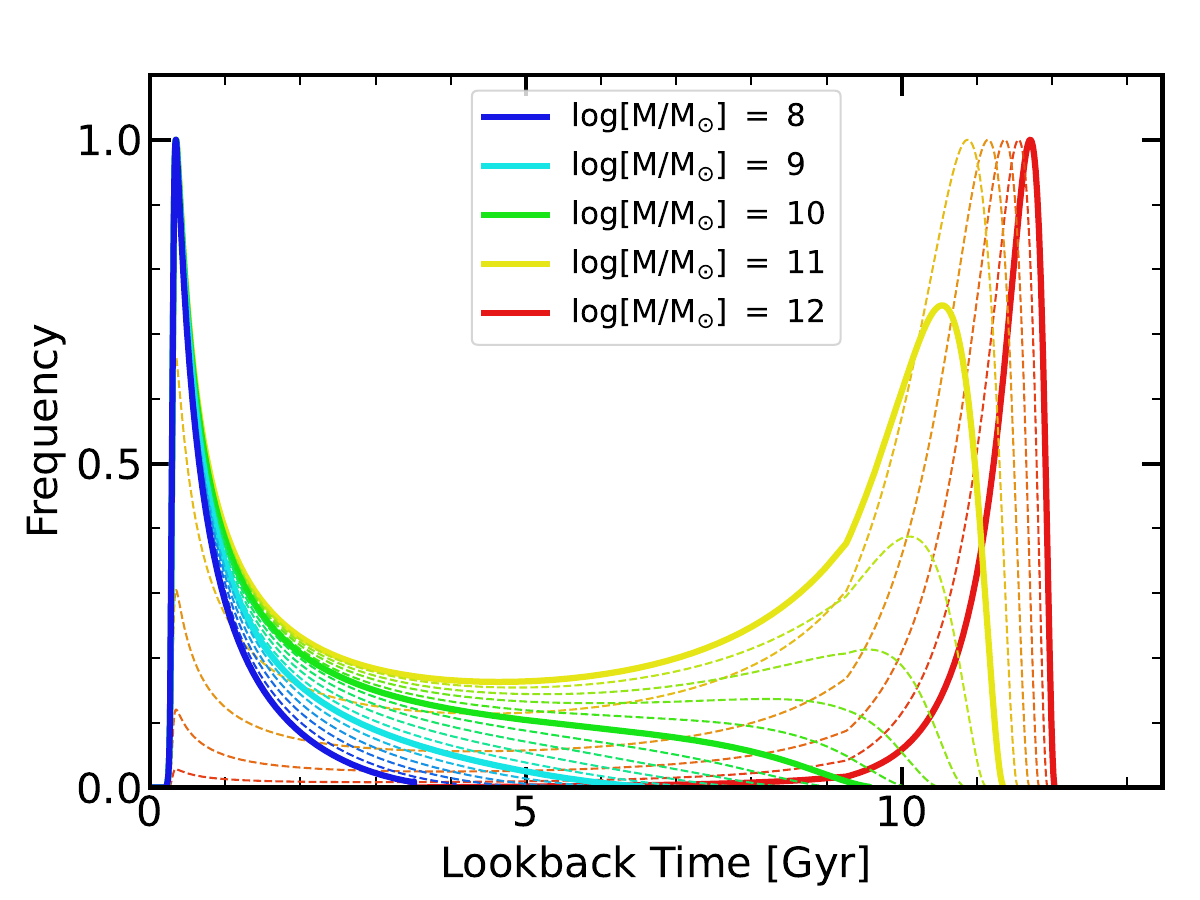}
    \caption
    {The SPAD along the galaxy stellar-mass sequence at $z = 0$, derived from \Cref{fig10} following \citetalias{Childress+2014MNRAS.445.1898C}.
    The color coding matches \Cref{fig9}. 
    Low-mass galaxies ($\log\mathrm{[M/M_{\odot}]} = 8 - 9$) predominantly host young SN Ia progenitors, whereas the most massive systems ($\log\mathrm{[M/M_{\odot}]} = 12$) preferentially host old progenitors.
    Intermediate-mass galaxies ($\log\mathrm{[M/M_{\odot}]} = 10 - 11$) exhibit a bimodal SPAD, containing both young and old progenitor populations.}
    \label{fig11}
\end{figure}

We briefly introduce the framework of \citetalias{Childress+2014MNRAS.445.1898C}.
\Cref{fig9} presents the observationally motivated SFHs across galaxy mass bins at $z = 0$.
The SFHs are unimodal and strongly skewed, varying systematically with mass. 
Low-mass galaxies form stars at later epochs and sustain star formation to the present, whereas high-mass galaxies build up rapidly at early times and subsequently quench.\footnote{\label{footnote2}The SFHs of low-mass galaxies adopted here underrepresent the contribution of old stellar populations \citep{Massana+2022MNRAS.513L..40M}, likely leading to an overestimation of the age difference between low- and high-mass hosts. Addressing this limitation will require future work adopting more realistic SFHs for low-mass galaxies that explicitly incorporate old stellar components.}
\Cref{fig10} shows that convolving these SFHs with the adopted delay-time distribution (DTD) formalism (\Crefapp{appA.3.sfh&spad}) yields the corresponding SPADs.
Low-mass galaxies ($\log\mathrm{[M/M_{\odot}]} = 8 - 9$) are dominated by young progenitors, with their contribution amplified near the DTD peak.
In contrast, the highest-mass bin ($\log\mathrm{[M/M_{\odot}]} = 12$) is dominated by old progenitors and carries little weight near the DTD peak.
Intermediate-mass galaxies ($\log\mathrm{[M/M_{\odot}]} = 10 - 11$) remain largely old but retain modest recent star formation (RSF), producing bimodal (young + old) SPADs.
\Cref{fig11} summarizes the SPADs across mass bins, showing that SPADs depend on galaxy mass as strongly as SFHs.

\begin{figure}
    \centering
    \includegraphics[width=1\linewidth]{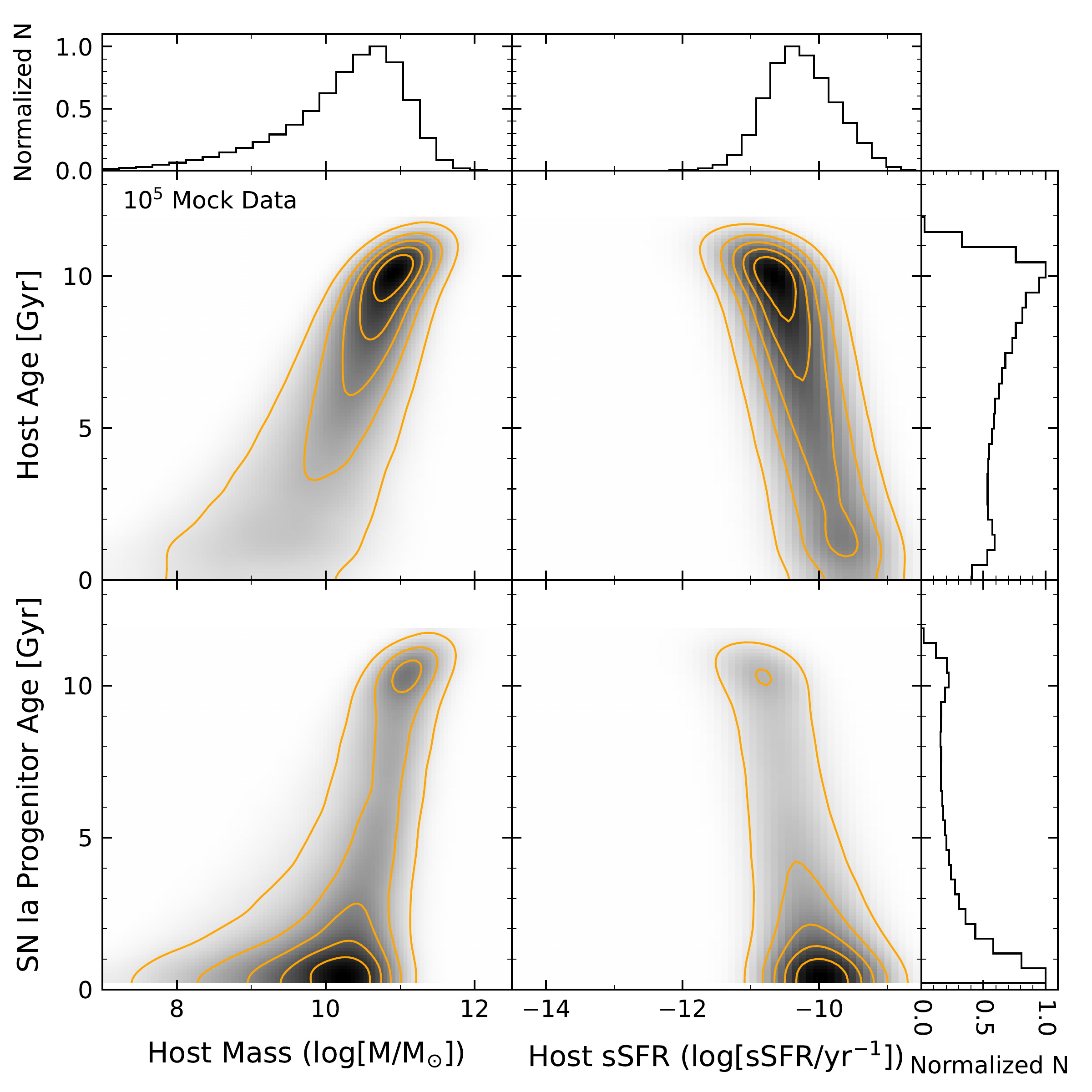}
    \caption
    {Density maps of stellar population age (top row) and SN Ia progenitor age (bottom row) versus host-galaxy stellar mass (left column) and specific star formation rate (sSFR; right column), derived from Monte Carlo realizations of $10^5$ mock galaxies at $z = 0$. 
    The marginal histogram along each axis shows the corresponding projected distribution.
    The population-age distribution is mildly bimodal, with a larger fraction in older hosts; its dominant peak coincides with the old stellar-population component and with the peak of the host-mass distribution at $\log\mathrm{[M/M_{\odot}]} \sim 10.7$.
    In contrast, the progenitor-age distribution is bimodal, with peaks at young and old ages and a substantially larger contribution from the young-progenitor component.}
    \label{fig12}
\end{figure}

We then perform Monte Carlo (MC) simulations\footnote{In these simulations, which follow the prescriptions of \citetalias{Childress+2014MNRAS.445.1898C}, the dominant sources of uncertainties enter primarily through the estimates of galaxy stellar mass and SFR. The stellar masses are tied mainly to the mass estimates based on \citet{Bruzual&Charlot2003MNRAS.344.1000B}, whereas the SFRs are inferred from empirical emission-line relations parameterized as a function of redshift \citep{Zahid+2012ApJ...757...54Z}. Since the host-mass distribution is imposed separately at each redshift, i.e., look-back time, additional uncertainties in the stellar population modeling have only a limited impact on the simulated host-galaxy populations.} to generate $10^5$ mock galaxies whose masses follow the observed SN Ia host mass function (Appendices~\ref{appA.1.anchor} and \ref{appA.2.host_mass}) and to assign their population- and progenitor-ages using quantities in \Cref{fig9,fig10,fig11}.
\Cref{fig12} shows the resulting density maps of population age (top row) and progenitor age (bottom row) versus host-galaxy mass (left column) and sSFR (right column) at $z = 0$.
The population-age distribution peaks at $\sim$\,10~Gyr corresponding to $\log\mathrm{[M/M_{\odot}]} \sim 10.7$ and $\log\mathrm{[sSFR/yr^{-1}]} \sim -10.4$, with a smooth tail toward lower masses and younger ages.
By contrast, the progenitor-age distribution is clearly bimodal, comprising a dominant low-mass, young component and a secondary massive, old component.
The old component aligns with the population-age peak, consistent with the predominance of old progenitors in old hosts, whereas the young component peaks near $\sim$\,0.3~Gyr (i.e., the DTD peak), outnumbering the old progenitors.
Because intermediate-mass hosts ($\log\mathrm{[M/M_{\odot}]} = 10 - 11$) both sit near the maximum of the SN Ia host mass function and exhibit the strongest bimodal SPADs (see \Cref{fig11}), they contribute disproportionately to the young peak despite their older population ages.
Although lower-mass galaxies also favor young progenitors, their overall contribution is reduced by their lower abundance in the host mass function.

\begin{figure}
    \centering
    \includegraphics[width=1\linewidth]{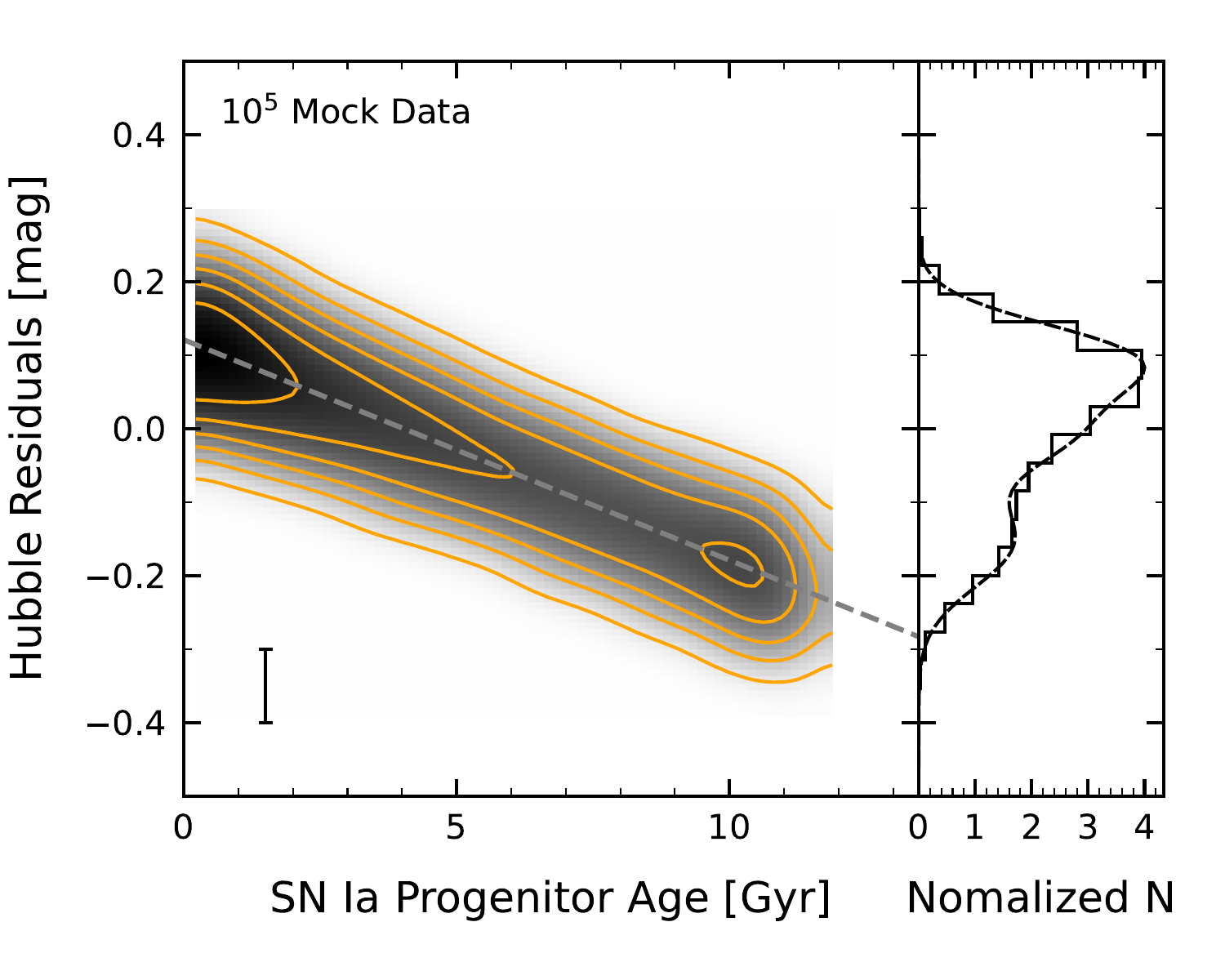}
    \caption
    {Density map of the simulated relation between SN Ia progenitor age and HR. 
    The MC simulation generates $10^5$ mock SNe Ia, adopting a linear progenitor-age--HR relation with slope $-0.030\,\mathrm{mag/Gyr}$ (gray dashed line) and an observational scatter of 0.05~mag (error bar at bottom left).
    The mean HR of the $10^5$ SNe Ia is set to zero. 
    The HR distribution is shown in the right panel (solid line). 
    A Gaussian mixture-model fit (dashed line) indicates a clear bimodality, with distinct peaks corresponding to young and old progenitors.}
    \label{fig13}
\end{figure}

\begin{figure*}
    \centering
    \includegraphics[width=0.61\linewidth]{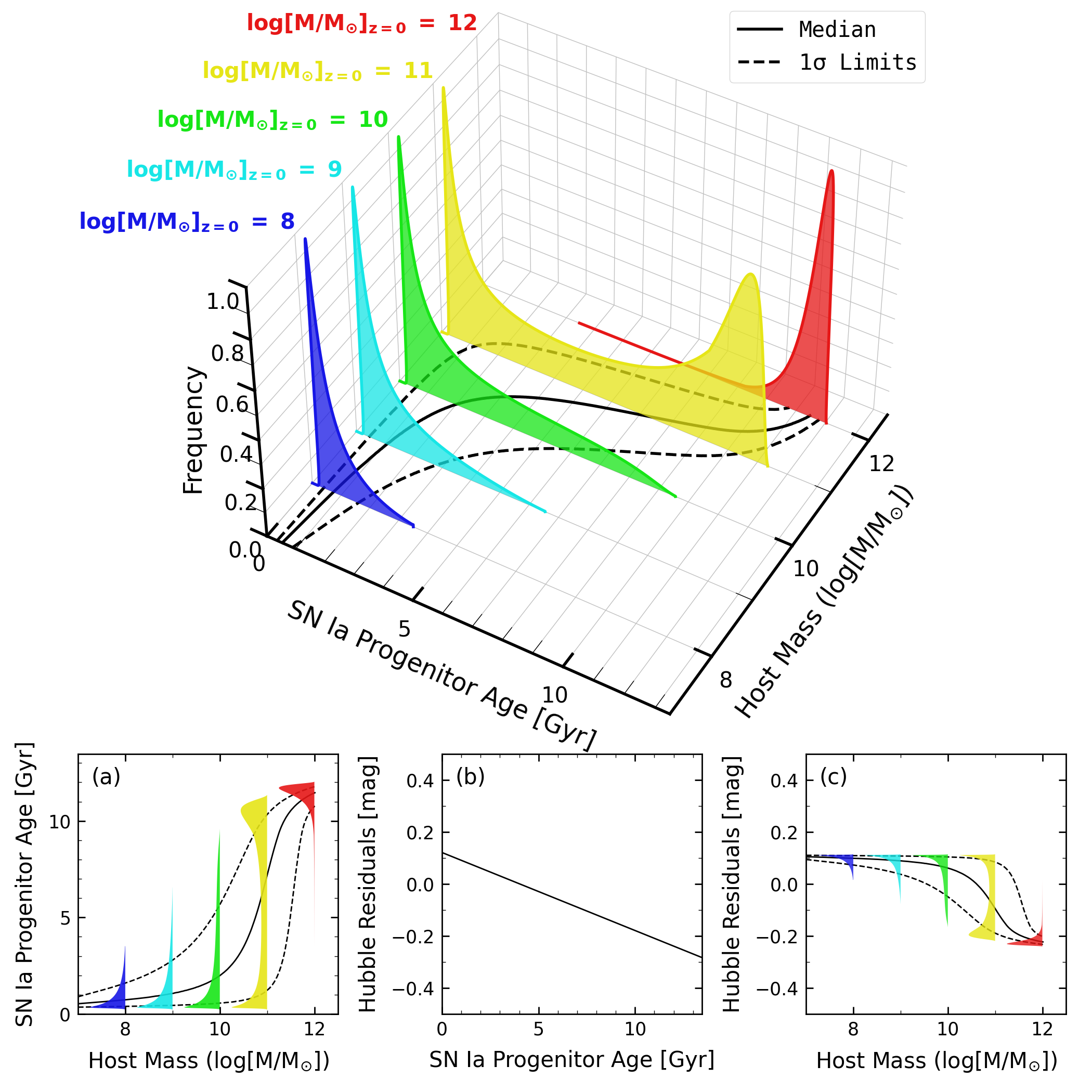}
    \caption
    {A schematic diagram illustrating how the interplay between progenitor age and host-galaxy properties produces the host mass step in HRs.
    \textit{Top:} A three-dimensional cube of host stellar mass, progenitor age, and event frequency at $z = 0$, constructed from the SPADs along the galaxy mass sequence (\Cref{fig11}).
    Intermediate-mass galaxies ($\log\mathrm{[M/M_{\odot}]} = 10 - 11$) exhibit strongly bimodal SPADs, hosting both young and old progenitors, whereas low-mass ($\log\mathrm{[M/M_{\odot}]} = 8 - 9$) and high-mass ($\log\mathrm{[M/M_{\odot}]} = 12$) galaxies show narrower SPADs dominated by young and old progenitors, respectively.
    This distribution leads to the nonlinearity of the progenitor-age-host-mass relation, as shown by the median (solid line) and $1\sigma$ bounds (dashed lines) on the $xy$-plane.
    \textit{Bottom:} How this nonlinearity generates a mass step.
    ($a$) Projection of the cube onto the $xy$-plane.
    The host-mass-mean progenitor-age relation is S-shaped and strongly nonlinear.
    ($b$) The empirical linear progenitor-age--HR relation adopted in this work (\Cref{fig13}).
    ($c$) The resulting schematic host mass step obtained by convolving ($a$) with ($b$).
    The rapid transition in mean progenitor age at $\log\mathrm{[M/M_{\odot}]} = 10 - 11$ in panel ($a$), through the linear relation in panel ($b$), yields the observed HR mass-step in panel ($c$).}
    \label{fig14}
\end{figure*}

\begin{figure*}
    \centering
    \includegraphics[width=0.7\linewidth]{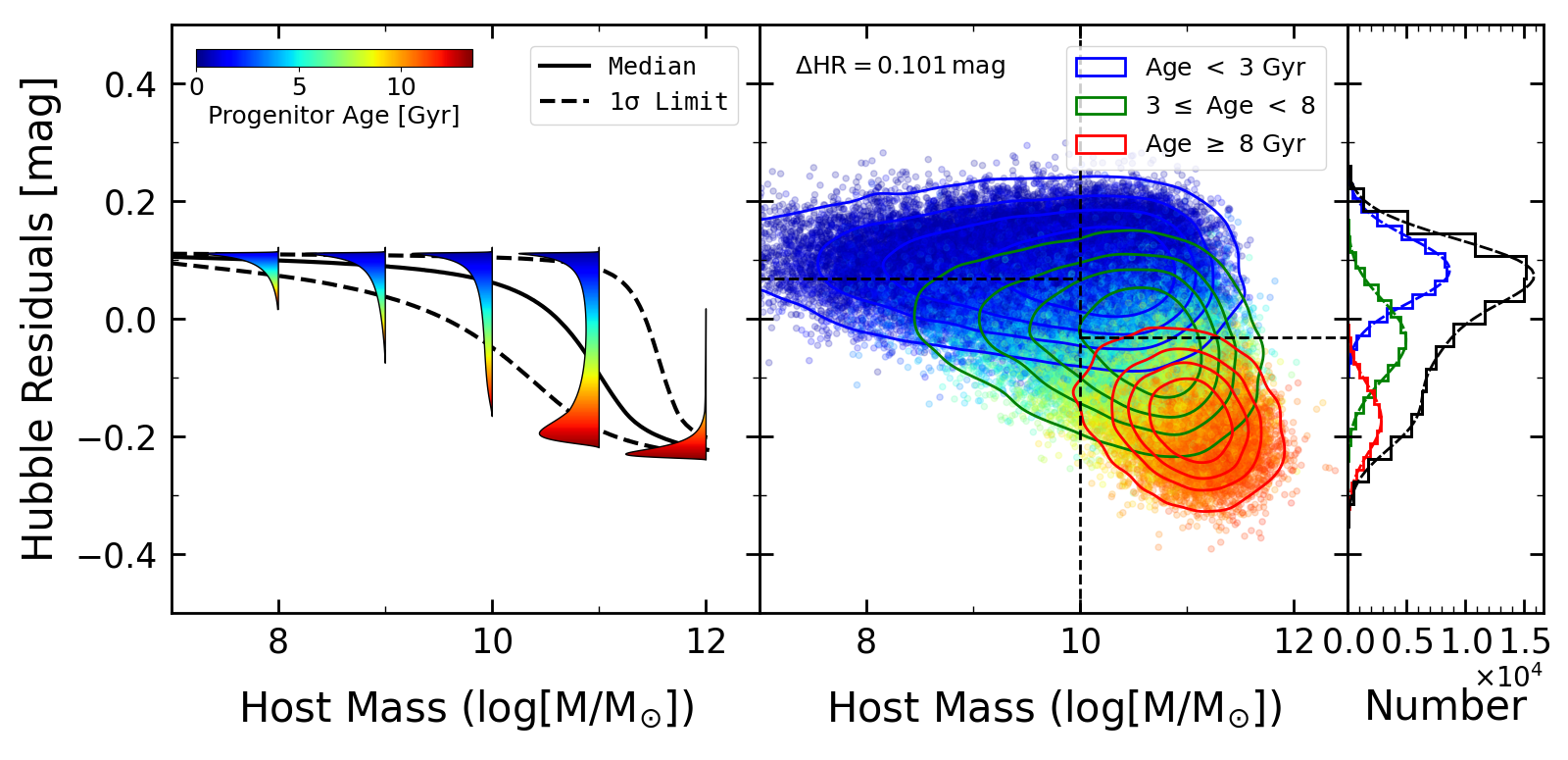}
    \caption
    {A simulation of the host-galaxy mass step in HRs.
    \textbf{\textit{Left:}} The same as \Cref{fig14}($c$), but with each SPAD colored by a spectrum encoding the SN Ia progenitor age, mapped to the corresponding HR.
    \textbf{\textit{Middle:}} Host-mass step simulated with $10^5$ mock host galaxies. 
    Points and contours, separated into young ($< 3 \, \text{Gyr}$; blue), intermediate ($3 \, \text{Gyr} \leq \text{age} < 8 \, \text{Gyr}$; green), and old ($\geq 8 \, \text{Gyr}$; red) progenitor-age subsets, show that an HR mass step emerges naturally from the SPAD structure along the galaxy mass sequence.
    Adopting a division at $\log\mathrm{[M/M_{\odot}]} = 10$ (vertical dashed line), the difference between the mean HRs on the low- and high-mass sides (horizontal dashed lines) corresponds to a mass step of $\Delta\mathrm{HR} = 0.101 \, \text{mag}$.
    \textbf{\textit{Right:}} Projected HR distributions for young (blue), intermediate-age (green), and old (red) progenitors. 
    Each subset yields a Gaussian HR distribution, whereas their superposition (dashed line) is bimodal and skewed toward fainter HRs because young progenitors dominate the number counts.}
    \label{fig15}
\end{figure*}

\begin{figure*}
    \centering
    \includegraphics[width=0.61\linewidth]{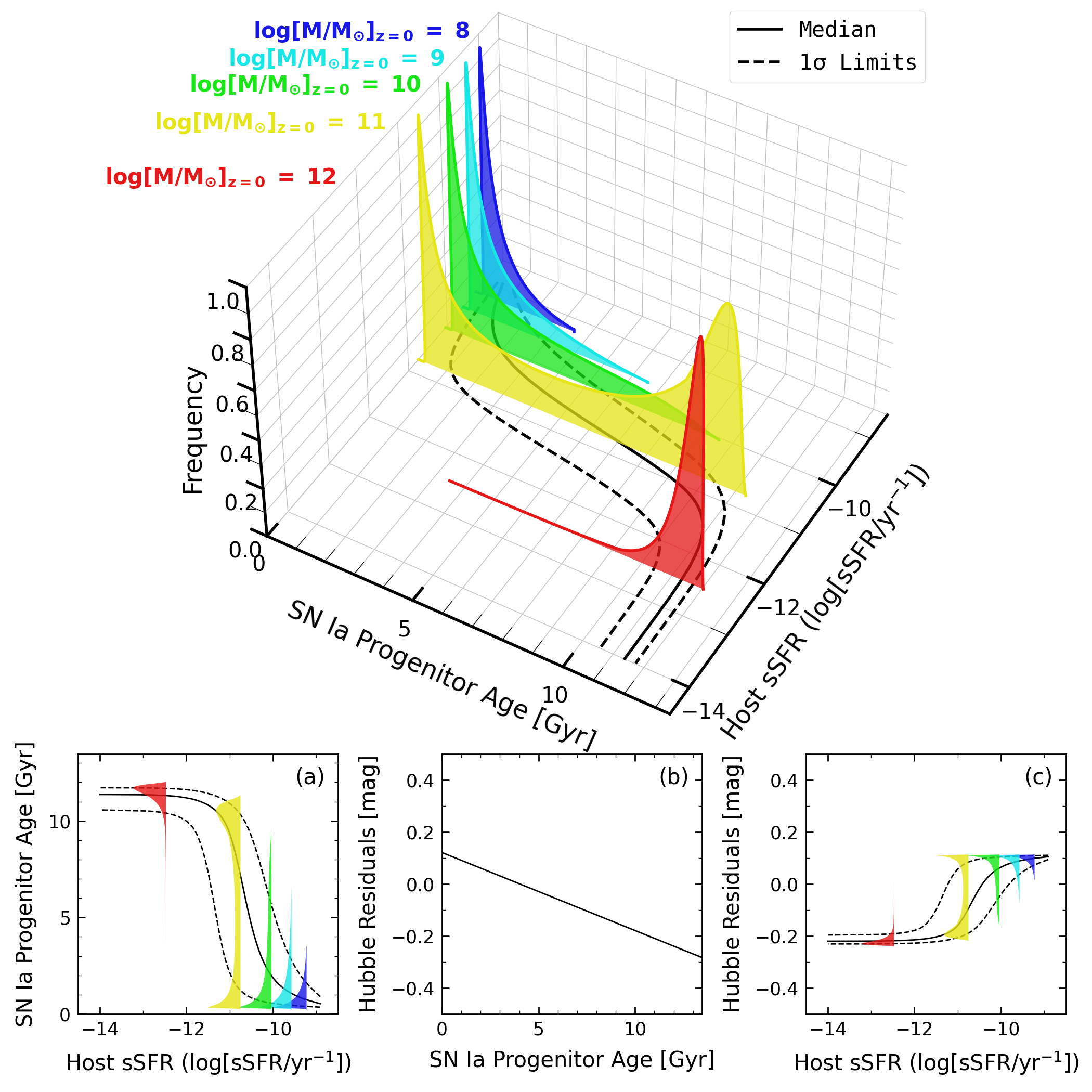}
    \caption
    {Closely analogous to \Cref{fig14}, but illustrating how the interplay between progenitor age and host sSFR produces the host sSFR-step in HRs.}
    \label{fig16}
\end{figure*}

\begin{figure*}
    \centering
    \includegraphics[width=0.7\linewidth]{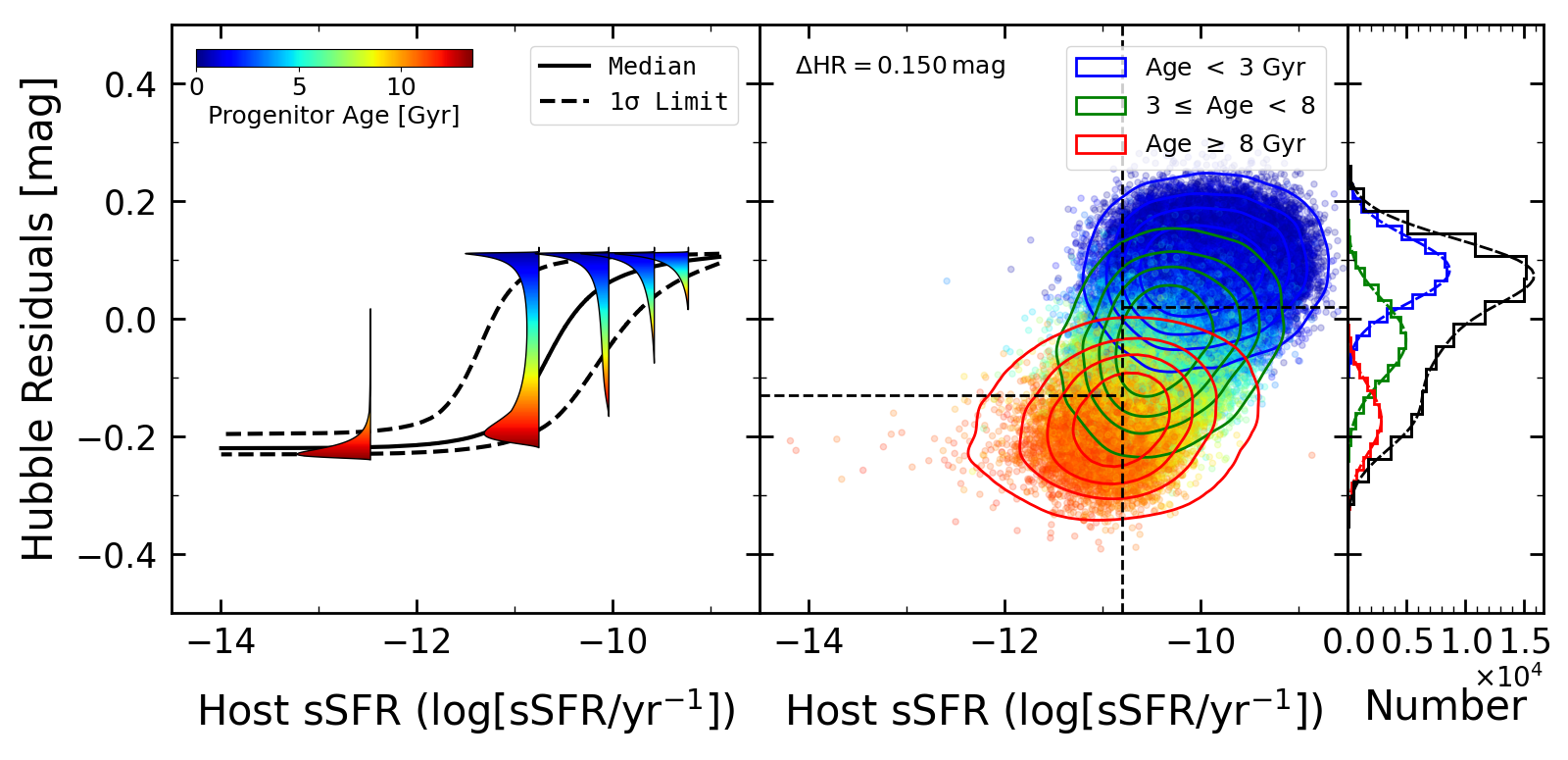}
    \caption
    {Closely analogous to \Cref{fig15}, but for the host sSFR step in HRs. 
    The left panel is the same as \Cref{fig16}($c$), but with each SPAD colored by a spectrum encoding the SN Ia progenitor age, mapped to the corresponding HR.  
    The sSFR step at $\log\mathrm{[sSFR / yr^{-1}]} = -10.8$ (the threshold identified by \citet{Rigault+2020A&A...644A.176R}) is measured to be $\Delta\mathrm{HR} = 0.150 \, \text{mag}$.}
    \label{fig17}
\end{figure*}

\begin{figure*}
    \centering
    \includegraphics[width=0.7\linewidth]{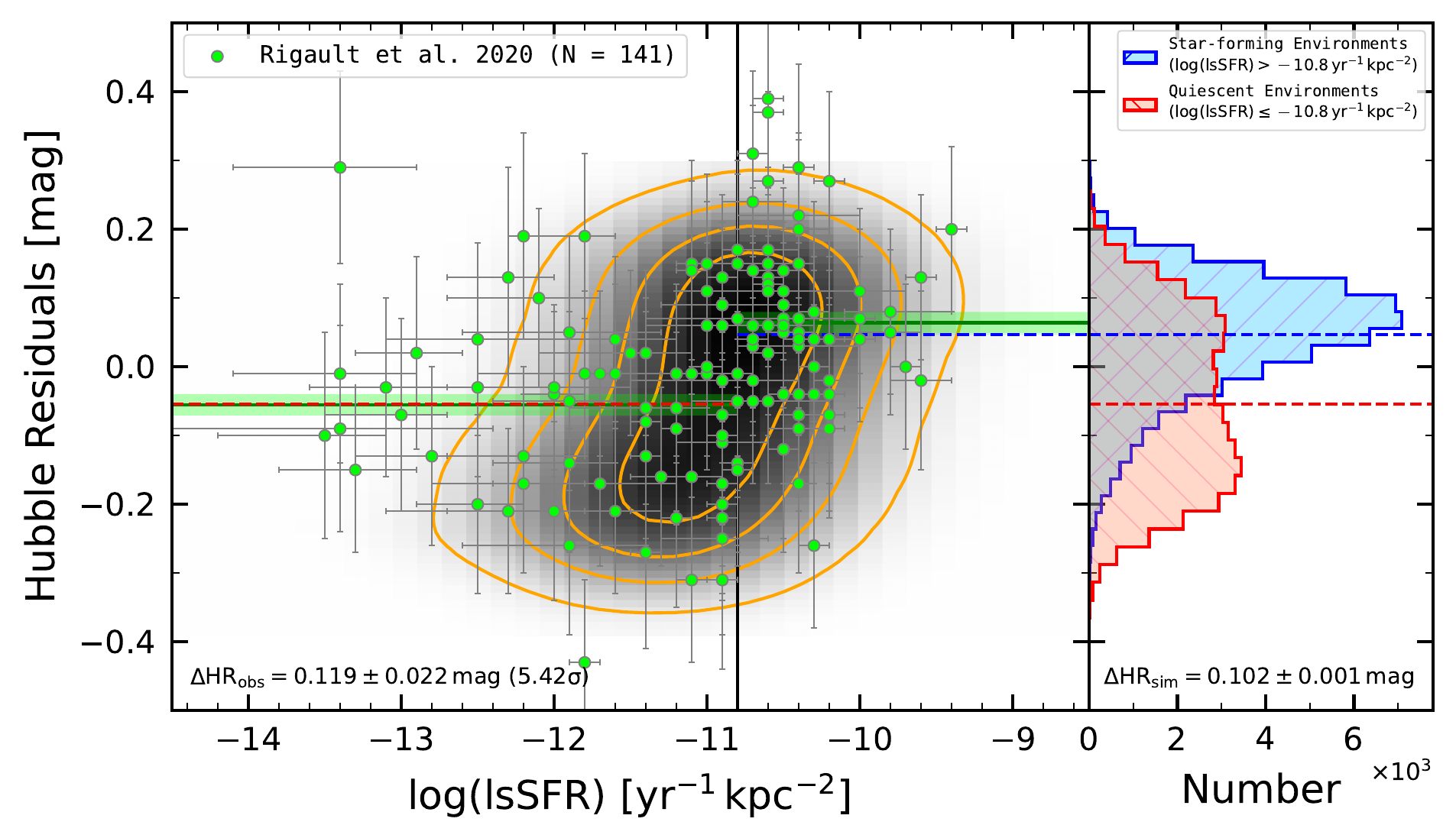}
    \caption
    {Comparison of the observed and simulated lsSFR steps.
    \textbf{\textit{Left:}} The lsSFR step measured by \citet{Rigault+2020A&A...644A.176R} is shown as green points with gray error bars.
    The density map (grayscale, with orange contours) is derived from an MC realization of $10^5$ mock hosts using the global-to-local sSFR conversion described in \Crefapp{appA.5.ssfr}.
    Locally star-forming and quiescent environments are separated by the threshold $\log(\mathrm{lsSFR}) = -10.8 \, \mathrm{yr^{-1} \, kpc^{-2}}$ (black vertical line).
    The mean HRs in the locally star-forming and quiescent regimes are shown as green horizontal lines, with $1\sigma$ ranges indicated by green shaded bands.
    The lsSFR-step amplitude, computed via an error-weighted Welch's two-sample \textit{t}-test \citep{Welch1947Biometrika.34.28} using the $1\sigma$ HR uncertainties of the observed data, is reported at bottom left ($\Delta\mathrm{HR_{obs}}$).
    The simulations reproduce the overall distribution, although the observed sample is skewed toward lower-sSFR hosts.
    This discrepancy may reflect residual star formation in old galaxies (e.g., triggered by recent tidal interactions), which would enhance the SN Ia counts at low sSFR relative to our model.
    \textbf{\textit{Right:}} HR distributions for the $10^5$ mock data in locally star-forming (blue hatched histogram) and quiescent (red hatched histogram) environments.
    The corresponding mean HRs are marked by dashed lines.
    The simulated lsSFR-step amplitude is reported at the bottom ($\Delta\mathrm{HR_{sim}}$).
    Overall, the simulated lsSFR step agrees well with the observations in both the distribution and step magnitude.}
    \label{fig18}
\end{figure*}

The final ingredient needed to reproduce SN Ia magnitude steps is a mapping between progenitor age and standardized SN Ia luminosity.
As described above, \citet{Lee+2020ApJ...903...22L} and \PaperI found that the inferred slope is statistically consistent between global and local population ages.
We therefore adopt the empirical relation inferred from host ages in \citet{Kang+2020ApJ...889....8K} and \PaperI.
Combining these datasets yields a conservative weighted-mean slope of $-0.030 \pm 0.004 \, \text{mag/Gyr}$ (see \PaperII).
In \Cref{fig13}, we apply this relation to the progenitor-age distribution (\Cref{fig12}) and use MC simulations of $10^5$ mock galaxies to construct the progenitor-age--HR relation (gray dashed line).
The bimodality in progenitor age (\Cref{fig12}) is transferred directly into a bimodal HR distribution.



\subsection{Reproducing SN Ia Magnitude Steps in Hubble Residuals}
\label{sec3.2.reproduce}

We now examine how the interplay between progenitor ages and host properties produces the HR magnitude steps.
We begin with the host mass step shown in \Cref{fig14}.
The top panel presents a 3D representation of host mass, progenitor age, and their frequency at $z = 0$, constructed from the SPADs along the galaxy mass sequence (\Cref{fig11}).
The projection onto the $xy$-plane defines the host mass--mean progenitor age relation, which is highly nonlinear.
The bottom panels illustrate how this nonlinearity generates the mass-step.
Panel ($a$) shows a face-on view of the 3D cube: low-mass and massive galaxies concentrate toward young and old progenitor ages, respectively, whereas intermediate-mass galaxies span a wide range of progenitor ages.
Panel ($b$) shows the assumed linear progenitor-age--HR relation.
Panel ($c$) then demonstrates that the convolution of panels ($a$) and ($b$) maps the sharp progenitor-age transition at $\log\mathrm{[M/M_{\odot}]} = 10 - 11$ onto the observed mass-step \citep[see also][]{Chung+2023ApJ...959...94C}.

\Cref{fig15} validates this picture using $10^5$ mock hosts from our MC simulations.
The left panel, analogous to \Cref{fig14}($c$), shows the HR distributions as a function of galaxy mass.
The middle panel presents the same simulated sample, separated into young ($< 3 \, \text{Gyr}$), intermediate (3\,$\sim$\,8 \, \text{Gyr}), and old ($> 8 \, \text{Gyr}$) subgroups.
The results show that the host mass-step emerges naturally from the SPAD structures along the galaxy mass sequence.
The right panel shows that each subgroup yields a Gaussian HR distribution, while their superposition becomes bimodal and skewed toward fainter HRs because young progenitors dominate in number.
At $\log~[\rm{M/M_\odot}] = 10.0$, conventionally adopted as the threshold, we measure a step amplitude of $0.101~\text{mag}$, consistent with observations \citep[e.g.,][]{Sullivan+2010MNRAS.406..782S, Childress+2013ApJ...770..108C, Betoule+2014A&A...568A..22B, Rigault+2020A&A...644A.176R, Briday+2022A&A...657A..22B}.

We next consider the host sSFR-step, which is produced by the same underlying logic.
\Cref{fig16} shows that the host sSFR--mean progenitor age relation is again highly nonlinear and S-shaped.
Panel ($a$) shows that low- and high-sSFR galaxies preferentially occupy old and young progenitor ages, respectively, whereas intermediate-sSFR galaxies span a wide range of progenitor ages.
Panel ($c$) presents the product of the convolution of panels ($a$) and ($b$), demonstrating that a sharp progenitor-age transition at $\log\mathrm{sSFR} = -11 \sim -10$ generates the sSFR-step.
\Cref{fig17} presents the corresponding simulated sSFR-step using $10^5$ mock hosts, directly analogous to \Cref{fig15}.
The simulation again indicates that the transition arises from SPAD structures along the sSFR sequence.
At $\log\mathrm{sSFR} = -10.8$, we measure a step amplitude of 0.150~mag, comparable to the value reported by \citet{Rigault+2020A&A...644A.176R} for the local sSFR (lsSFR) step, $0.163\pm0.029\,\text{mag}$ at $\log\mathrm{lsSFR}=-10.8\,\mathrm{yr^{-1} kpc^{-2}}$.

Finally, \Cref{fig18} compares the simulated and observed lsSFR steps.
In the left panel, we compare $10^5$ mock SNe Ia from the MC simulation, after a global-to-local sSFR conversion (\Crefapp{appA.5.ssfr}), with the observed sample from \citet{Rigault+2020A&A...644A.176R}.
The simulations reproduce the overall distribution, but the observation shows a low-lsSFR tail at $\log\mathrm{lsSFR}\sim -13\,\mathrm{yr^{-1}\,kpc^{-2}}$.
This mismatch can be attributed to galaxy-to-galaxy variations in SFH among low-lsSFR host galaxies, together with scatter introduced by the global-to-local sSFR conversion in our model.
We divide both the simulated and observed samples into star-forming and quiescent environments at $\log\mathrm{lsSFR} = -10.8 \, \mathrm{yr^{-1} kpc^{-2}}$ (black vertical line).
Using Welch’s two-sample \textit{t}-test \citep{Welch1947Biometrika.34.28}, we measure an observed lsSFR step of $0.119 \pm 0.022 \, \mathrm{mag}$.
The right panel shows the HR distributions for $10^5$ mock SNe~Ia, analogous to the right panel of \Cref{fig17}, but with the sample divided into two lsSFR-defined subgroups.
Encouragingly, the resulting simulated lsSFR step, $0.102 \pm 0.001 \, \text{mag}$, is fully consistent with the observed value.



\section[Discussion]{Discussion}
\label{sec4.discussion}

We have shown that progenitor-age bias is the most likely underlying cause of the reported correlations between host properties and the HRs.
We have further demonstrated that both the host mass step and the sSFR step arise from the convolution of two relations: a nonlinear relation between progenitor age and host mass (or sSFR), and a linear relation between progenitor age and the standardized SN Ia magnitude.
Consequently, the observed SN Ia magnitude steps are not independent physical phenomena in their own right, but rather derivative effects arising from a single underlying relation: the systematic dependence of standardized SN Ia magnitude on progenitor age.

Since SNe Ia occur in galaxies spanning a wide range of stellar populations, their hosts naturally trace the underlying galaxy demographics.
In our simulation, this appears as the familiar blue cloud, green valley, and red sequence in age--mass--sSFR space, where these properties are strongly coupled \citep[e.g.,][]{Strateva+2001AJ....122.1861S, Kauffmann+2003MNRAS.341...54K}.
This demographic structure produces nonlinear relations between host-galaxy properties and progenitor age.
Low-mass, blue-cloud galaxies with ongoing star formation preferentially host young progenitors, whereas massive, red-sequence galaxies with quenched star formation preferentially host old progenitors.
Intermediate-mass galaxies in the green valley contain both young and old progenitors, forming a transitional population between these two regimes.
As a result, the age--mass and age--sSFR relations become nonlinear and S-shaped, reflecting the systematic change in stellar populations across galaxy types (see \Cref{fig14,fig16}).
In particular, the mapping from host-galaxy properties to progenitor age is closely linked to the strong bimodality of the SPAD in intermediate-mass, green-valley galaxies.
More broadly, similarly to a bimodal progenitor-age distribution (see \Cref{fig12,fig13}), the same galaxy demographic structure naturally produces a bimodal stellar-age distribution driven by the nonlinear age--mass relation \citep{Chung+2023ApJ...959...94C} across low- and high-mass galaxies \citep[e.g.,][]{Kauffmann+2003MNRAS.341...54K, Mateus+2006MNRAS.370..721M, Peng+2010ApJ...721..193P}.
Consequently, even if the dependence of SN Ia HR on progenitor age is intrinsically linear, its projection onto host-galaxy observables produces apparent magnitude steps (see \Cref{fig15,fig17}). 
The nonlinear age--mass (or age--sSFR) relation, together with the bimodal age distribution, further strengthens the amplitude and sharpness of these step-like features.

To estimate the impact of progenitor-age bias on cosmological inferences, it is necessary to compute the redshift-dependent magnitude correction, as described in \PaperII.
This correction is obtained by convolving the progenitor age versus HR slope ($\Delta\text{HR}/\Delta\text{age}$, hereafter the ``progenitor-age slope'') with the redshift evolution of the median progenitor age relative to $z = 0.0$.
The resulting cosmological effect therefore depends jointly on both the progenitor-age slope and the redshift evolution of progenitor age.
However, because the progenitor ages of SNe Ia cannot be measured directly, it is essential to evaluate how reliably the progenitor-age slope can be inferred from the measured stellar population ages of host galaxies (host ages).
Using the same methodology adopted for the sSFR simulations described above, we therefore investigate the relationship between SN Ia progenitor age and host age.
Our simulations show that the relation between host age and HR exhibits a non-negligible degree of nonlinearity, which may be visible in \Cref{fig1} (lower left panel).
Nevertheless, when a linear relation is assumed in the regression analysis, the inferred slope agrees with the input progenitor-age slope adopted in the simulations to within 5\%.
This result demonstrates that the progenitor-age slope can be robustly estimated from directly measured host ages, which is a highly encouraging outcome.

This result is obtained using the DTD proposed by \citetalias{Childress+2014MNRAS.445.1898C}, which represents an average DTD among those reported in the literature.
Recently, \citet{Wiseman+2026arXiv260113785W} argued that adopting a DTD with a much shorter prompt time and a different functional form ($\beta$) leads to a significantly weaker redshift evolution of progenitor age, reduced to approximately one third of that reported by \PaperII, and consequently to a negligible cosmological impact from progenitor-age bias.
Following the same methodology described above, we repeat the simulations using the DTD advocated by \citet{Wiseman+2026arXiv260113785W}.
In contrast to the case using the DTD adopted in \PaperII, which is also adopted in this work, we find that the progenitor-age slope inferred from the directly measured host-age slope becomes approximately three times larger than the host age slope itself.
This result arises naturally because, for a given difference in host age, the corresponding difference in progenitor age is substantially reduced.
Therefore, the reduction in the redshift evolution of progenitor age is accompanied by a compensating increase in the progenitor-age slope.
When these two effects are convolved to compute the redshift dependent magnitude correction, the final correction, and hence the resulting cosmological impact, remains largely unchanged.
While \citet{Wiseman+2026arXiv260113785W} correctly caution against conflating host age with progenitor age, they overlook the fact that the same effect that suppresses progenitor-age evolution simultaneously enhances the inferred progenitor-age slope.
A detailed discussion of these issues will be presented in our forthcoming counter rebuttal paper responding to \citet{Wiseman+2026arXiv260113785W}.

One of the predictions of our interpretation is that the amplitude of the host mass-step would mildly decrease with increasing redshift.
This behavior arises because the absolute ages of old progenitors in the SPAD rapidly decrease toward higher redshift, causing the overall age range to shrink and the distribution to evolve from bimodal toward unimodal (see \citetalias{Childress+2014MNRAS.445.1898C}).
This predicted redshift evolution of the mass step may qualitatively resemble what is observed empirically \citep{Rigault+2013A&A...560A..66R, Betoule+2014A&A...568A..22B, Childress+2014MNRAS.445.1898C, Wiseman+2026arXiv260113785W}, although a more systematic and quantitative comparison is still required, particularly with a more realistic SFH model for the low-mass galaxies as discussed in \Cref{footnote2}.
Similarly, our result also provides a natural interpretation for the recent finding by \citet{Toy+2025MNRAS.538..181T}, who reported that the mass step is suppressed in the outer regions of host galaxies.
This can be understood because the inner regions of late-type galaxies contain a mixture of old bulge populations and young disk populations, resulting in a wide span of stellar ages, whereas the outer regions are dominated primarily by young disk populations, leading to a much narrower age range and hence a reduced mass step.
A qualitatively similar difference in age range between the inner and outer regions is also expected in early-type host galaxies \citep{Koleva+2011MNRAS.417.1643K, Gomes+2016A&A...585A..92G, Gomes+2016A&A...588A..68G}.
Our model also predicts a mild redshift evolution in the distribution of the light-curve stretch parameter $x_1$.
This arises because, as shown in \Cref{fig3} (see also \citetalias{Lee+2022MNRAS.517.2697L}), low-stretch SNe Ia ($x_1 < -0.7$) are produced exclusively by old progenitors, whereas high-stretch SNe Ia can originate from both young and old progenitors.
This prediction is qualitatively consistent with the observed trends reported by Kim et al. (2026, submitted), while a more detailed quantitative analysis will be presented in a forthcoming paper.

Our study implies that applying the host mass step or sSFR step as independent correction terms in SN Ia luminosity standardization---rather than correcting directly for age---can bias cosmological inferences, because these steps are merely projected manifestations of the underlying age dependence.
One might argue that applying a host mass step correction already accounts for part of the host age effect.
Indeed, at a fixed redshift, this empirical correction can reduce some of the scatter in the HRs.
However, cosmological conclusions are driven primarily by systematic corrections that evolve with redshift, and it is precisely in this context that reliance on the host mass step alone becomes problematic in the absence of an explicit progenitor age correction.
The reason is straightforward: host mass cannot replace age.
These two quantities evolve very differently with redshift.
Within the redshift range most relevant to SN cosmology ($0 < z < 1$), the evolution of galaxy mass is relatively small or negligible, because the bulk of mass assembly was already completed by $z \approx 1 - 2$ \citep{Bell+2004ApJ...608..752B, Conroy+2007ApJ...668..826C}.
In contrast, the evolution of stellar populations in galaxies, including SN Ia progenitors, is substantial over the same redshift interval \citepalias{Childress+2014MNRAS.445.1898C, Lee+2022MNRAS.517.2697L, Son+2025MNRAS.544..975S}.
As discussed above, the redshift-dependent magnitude correction based on progenitor-age bias is relatively insensitive to the assumed DTD.
As already demonstrated in \PaperII, correcting SN Ia magnitudes using only the host mass step produces little change in the inferred cosmological model, particularly when all three cosmological probes (SNe Ia, BAO, and CMB) are considered together.
This arises because two host galaxies with the same stellar mass but at different redshifts---and therefore with different progenitor ages---would receive identical corrections in this approach.
By contrast, applying the progenitor-age bias correction results in dramatic shifts in the preferred cosmological model.
Therefore, it is essential to pursue the most direct SN-based cosmological tests that are free from concerns about progenitor-age bias and the resulting overcorrection in the luminosity standardization process.
To achieve this, one must perform an ``evolution-free'' test that uses only SNe Ia hosted by galaxies with uniformly young stellar populations across the entire redshift range.
By construction, such a sample removes the possibility of luminosity evolution effects and provides a clean cosmological test that is intrinsically insensitive to progenitor-age bias.
We therefore urge the community to pursue future SN cosmology studies along these lines.



\section*{Acknowledgements}

We thank the referee for a number of helpful comments and suggestions.
We acknowledge support from the National Research Foundation of Korea to the Center for Galaxy Evolution Research (RS-2022-NR070872, RS-2022-NR070525).
S.-J.Y. acknowledges support from the Mid-career Researcher Program (RS-2024-00344283) through Korea's NRF funded by the Ministry of Science and ICT.
Y.-L.K. was supported by the Lee Wonchul Fellowship, funded through the BK21 Fostering Outstanding Universities for Research (FOUR) Program (grant No. 4120200513819).



\section*{Data Availability}
 
There are no new data associated with this article.




\bibliographystyle{mnras}

\bibliography{references} 




\appendix



\section[Monte Carlo Simulation of SN Ia Host Properties]{Monte Carlo Simulation of \\ SN Ia Host Properties}
\label{appA.simulation}

\begin{figure*}
    \centering
    \includegraphics[width=\linewidth]{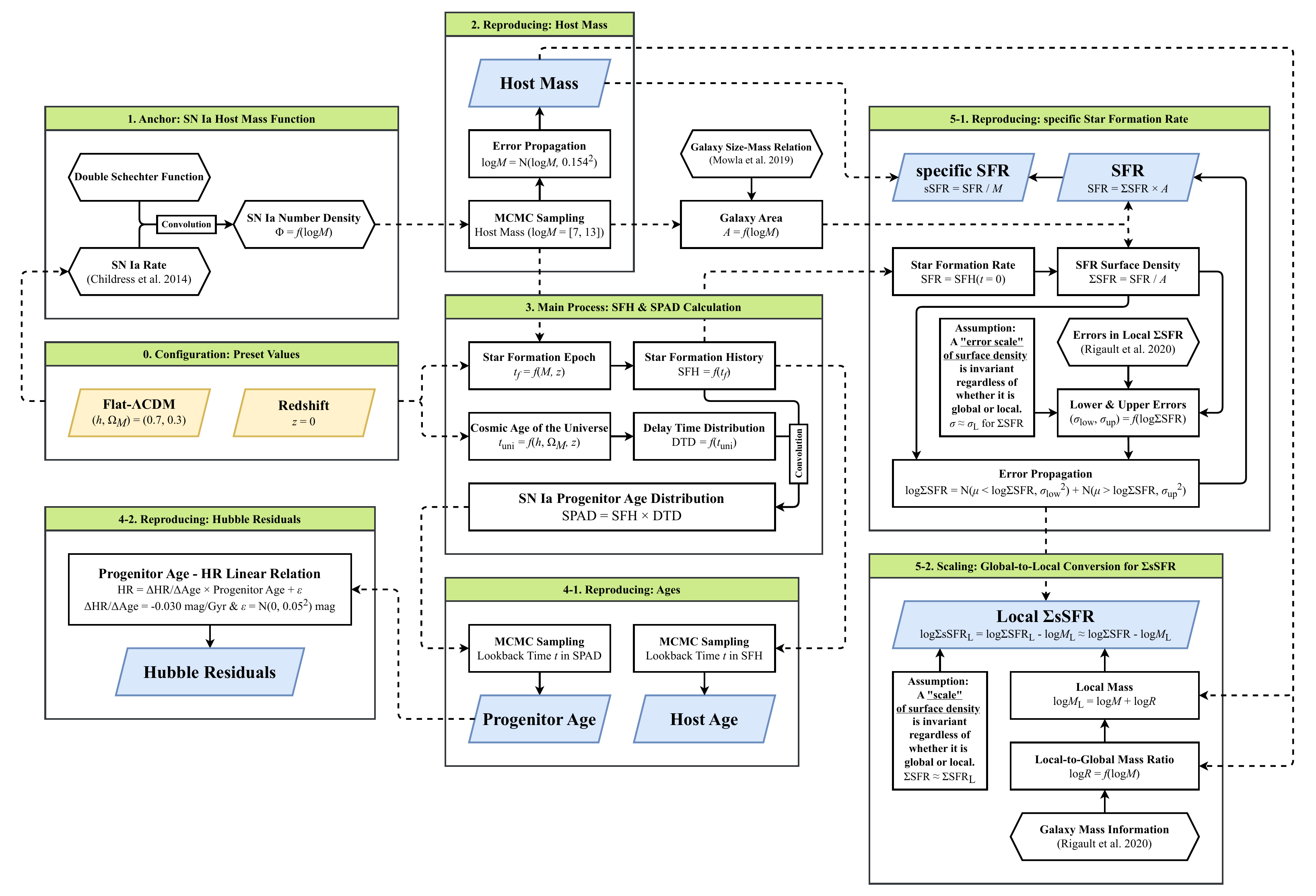}
    \caption
    {Flowchart of Monte Carlo (MC) simulation to reproduce SN Ia host properties.
    The MC simulation consists of several processes (green entitled boxes), each of which integrates its own operations (briefly described in symbols).
    The orange parallelograms in the 0th process represent the starting points, where computing environments are configured.
    The blue parallelograms in the other processes indicate output that lead to SN Ia host properties.
    Flows between operations within the same process are connected by solid arrows, and flows between operations integrated into different individual processes are connected by dashed arrows.
    The descriptions for each operation and process are referred to \Cref{appA.simulation} in detail.}
    \label{figA1}
\end{figure*}

The overall MC simulation procedure is illustrated in \Cref{figA1}.
The simulation assumes a flat $\Lambda$CDM cosmology with $H_0 = 70\,\mathrm{km\,s^{-1}\,Mpc^{-1}}$ and $\Omega_M = 0.3$.
The following subsections correspond to the processes shown in \Cref{figA1}, each labeled with the same subtitle.

\subsection{SN Ia Host Mass Function}
\label{appA.1.anchor}

We derive the host-galaxy stellar mass function of SNe Ia at $z = 0$, which reflects the SN Ia number density, by convolving the double-Schechter function with the SN Ia rate.
The SN Ia rate is derived as a function of host mass from the main procedure for computing the galaxy SFH and SPAD (see \Crefapp{appA.3.sfh&spad}).

To simulate the host-mass distributions of SNe Ia, we adopt the galaxy luminosity function parameterized by a Schechter function \citep{Schechter1976ApJ...203..297S}.
In modeling the galaxy stellar mass function, recent studies have employed a double Schechter function, consisting of two single-Schechter components that represent star-forming and quiescent galaxies \citep[e.g.,][]{Baldry+2008MNRAS.388..945B, Baldry+2012MNRAS.421..621B, Peng+2010ApJ...721..193P, Muzzin+2013ApJ...777...18M, Tomczak+2014ApJ...783...85T}.
Accordingly, we use the general form of the double Schechter function given below:
\begin{equation}
    \Phi \times \mathrm{d}M= \left[\Phi(\alpha_1, M^*_1, \Phi^*_1) + \Phi(\alpha_2, M^*_2, \Phi^*_2)\right] \times \mathrm{d}M.
\label{eqA1}
\end{equation}
If both components of double Schechter function share the same $M^*$, it can be expressed in a simplified logarithmic form,
\begin{equation}
\begin{split}
    \Phi \times \mathrm{d}(\log M) = & \ln (10) \times \exp\left(-10^\mathcal{M}\right) \\
                                     & \times \left[\Phi_1^* \left(10^\mathcal{M}\right)^{\alpha_1 + 1} + \Phi_2^* \left(10^\mathcal{M}\right)^{\alpha_2 + 1} \right] \mathrm{d}(\log M),
\end{split}
\label{eqA2}
\end{equation}
where $\mathcal{M} \equiv \log M - \log M^*$.
In our simulation, we adopt the parameter sets from \citetalias{Childress+2014MNRAS.445.1898C}, $(\alpha_1, \log M^*_1, \Phi^*_1) = (-1.42, 11.12, 1.39 \times 10^{-3})$ and $(\alpha_2, \log M^*_2, \Phi^*_2) = (-0.45, 10.66, 6.57 \times 10^{-3})$, for star-forming and quiescent galaxies, respectively.



\subsection{Host Mass}
\label{appA.2.host_mass}

We generate a set of stellar masses for SN Ia host galaxies based on the SN Ia number density using MCMC sampling.
The host-mass distribution is limited to $\log\mathrm{[M/M_{\odot}]} = [7.0,\, 13.0]$.
To reproduce observational uncertainties, we apply a Gaussian error of $\sigma_{\log\mathrm{[M/M_{\odot}]}} = 0.154$, corresponding to the mean stellar-mass uncertainty in the \citet{Rigault+2020A&A...644A.176R} dataset.
Minor variations in stellar mass error do not affect our main results.



\subsection{SFH \& SPAD Calculation}
\label{appA.3.sfh&spad}

The generated host-mass distribution is used to compute the galaxy SFH and SPAD before applying error propagation.
We employ the prescriptions of \citetalias{Childress+2014MNRAS.445.1898C} to calculate the SFHs and SPADs of SN Ia host galaxies as a function of stellar mass and redshift (see Appendix in \citetalias{Childress+2014MNRAS.445.1898C}).
The star formation epoch ($t_f$) that yields the resulting stellar mass at $z = 0$ is interpolated from the evolutionary sequences of galaxy mass using the same procedure as in \citetalias{Childress+2014MNRAS.445.1898C}.
Each SFH is integrated from $t_f$ to $z = 0$ and convolved with the DTD to produce the SPAD.

To incorporate the DTD ($\phi$), which represents the theoretical SN Ia rate in a galaxy as a function of time ($t$), we adopt the smooth functional form introduced by \citetalias{Childress+2014MNRAS.445.1898C}, as below:
\begin{equation}
    \phi\left(t\right) \propto \frac{\left(t / t_p\right)^{\alpha}}{\left(t / t_p\right)^{\alpha - s} + 1} \, .
    \label{eqA3}
\end{equation}
The prompt timescale is denoted by $t_p$. 
We adopt $t_p = 0.3 \ \mathrm{Gyr}$ with $s = -1$ and $\alpha = 20$, following \citet{Kang+2020ApJ...889....8K}.
The DTD is normalized as a PDF over the time domain $[0, t_{\rm{uni}}]$, where $t_{\rm{uni}}$ is the age of the Universe at $z = 0$.

The SPAD ($\mathcal{P}$) is obtained as the product of the galaxy SFH ($\psi$) and the DTD ($\phi$), as given by
\begin{equation}
    \mathcal{P}(\tau; t) = \psi(t - \tau) \phi(\tau) \, .
    \label{eqA4}
\end{equation}
Thus, the SN Ia rate in a galaxy ($\mathcal{R}$) is given by the integral of its SPAD, as expressed below:
\begin{equation}
    \mathcal{R} = \int_{0}^{t - t_f} \mathcal{P}(\tau; t) \, d\tau \, .
    \label{eqA5}
\end{equation}



\subsection{Ages \& Hubble Residuals}
\label{appA.4.ages&hrs}

We predict the ages of the SN Ia host galaxy and its progenitor from the SFH and SPAD, respectively.
For each host and progenitor age, we generate 100 candidates following the corresponding SFH and SPAD posteriors in the MCMC sampling, and one candidate is randomly selected.
The HR of each SN Ia is then computed using the empirical linear relation with progenitor age, adopting a conservative mean slope of $\Delta \text{HR} / \Delta \text{age} = -0.030\,\mathrm{mag/Gyr}$ (see \Cref{sec3.nonlinearity}).
The HR distribution is shifted to have a mean of zero, and Gaussian errors of 0.05~mag are added to each HR value.



\subsection{sSFR \& Global-to-Local Conversion}
\label{appA.5.ssfr}

We derive the global SFR at $z = 0$ from each galaxy's SFH and reproduce the associated observational uncertainties.
\citet{Rigault+2020A&A...644A.176R} provide reliable stellar-mass and SFR measurements for local SN Ia environments.
To apply these local SFR uncertainties ($\left[\mathrm{M_{\odot}/yr/kpc^2}\right]$) to the global SFR ($\left[\mathrm{M_{\odot}/yr}\right]$), we assume that the error scale of the SFR surface density, $\Sigma\mathrm{SFR}$ ($\left[\mathrm{M_{\odot}/yr/kpc^2}\right]$), is invariant for both global and local measurements.
The lower ($\sigma_{\rm{low}}$) and upper ($\sigma_{\rm{up}}$) errors of the local SFR measurements are fitted as quadratic functions of $\log\Sigma\rm{SFR}$.
Because of the logarithmic scale, $\sigma_{\rm{up}}$ tends to be smaller than $\sigma_{\rm{low}}$.
Values of $\log\Sigma\rm{SFR}$ outside the range of the \citet{Rigault+2020A&A...644A.176R} dataset are assigned the edge values of the fitting functions.
Consequently, the error propagation of $\log\Sigma\rm{SFR}$ for both global and local measurements is represented by a split-normal distribution, as expressed below:
\begin{equation}
    \log\Sigma\mathrm{SFR} = \mathrm{N}(\mu <    \log\Sigma\mathrm{SFR}, {\sigma_{\mathrm{low}}}^2) +
                             \mathrm{N}(\mu \geq \log\Sigma\mathrm{SFR}, {\sigma_{\mathrm{up}}}^2) \, .
\end{equation}
Since local SFR measurements are expressed as surface densities, we divide the global SFR of each host galaxy by its predicted area using observed galaxy size-mass relation \citep{Mowla+2019ApJ...880...57M} to ensure consistency between the two.

To reproduce local sSFR (lsSFR) from global sSFR---thereby mimicking SN Ia environments---we adopt an approach analogous to the sSFR reproduction method and assume that the scale of $\log\Sigma\rm{SFR}$ is invariant between the host galaxy as a whole and local SN Ia site.
Because lsSFR measurements require accounting for the stellar mass enclosed within a fixed observational aperture, we model the local-to-global mass ratio, $\log R$, using the host galaxy mass information from \citet{Rigault+2020A&A...644A.176R}, who employed a 1kpc aperture.
This procedure enables us to reproduce lsSFR values from the host-galaxy sSFR values generated by the MC simulations.



\bsp    
\label{lastpage}
\end{document}